# A Spherical Harmonic Analysis of Redshift Space

A.F. Heavens and A.N. Taylor
*Institute for Astronomy, University of Edinburgh, Royal Observatory, Blackford Hill, Edinburgh, U.K.*

**ABSTRACT**

We re-examine the effects of redshift space distortion in all–sky galaxy redshift surveys in the formalism of spherical harmonics. This natural decomposition of the density field into radial and angular eigenfunctions of the Laplacian operator complements both the spherical symmetry of the survey geometry and the dynamical basis for the redshift distortion.

Within this framework we show how one can treat both the large–scale linear effects, and the small-scale nonlinear clustering, exactly to first order. We show, contrary to earlier claims, that the redshifted density field is no longer homogeneous, as well as being anisotropic.

We construct a likelihood function for each mode in the decomposition, based on the random Gaussian field hypothesis. This function finds its maximum when the underlying field is homogeneous and isotropic, and has the correct amplitude at each mode. As the level of distortion in the field is governed by the cosmological density parameter, via $\beta \equiv \Omega_0^{0.6}/b$, where $b$ is the galaxy bias parameter, strong limits can be placed on cosmological models.

The method also allows in principle a determination of the power spectrum of perturbations, requiring no assumptions beyond that of linear theory. The method therefore offers significant advantages over Fourier techniques when dealing with all-sky surveys.

We apply our likelihood analysis to both simulated data, and real data, using the IRAS 1.2-Jy galaxy catalogue, for which we find a maximum likelihood $\beta \simeq 1.1 \pm 0.3$, and a real-space fluctuation amplitude corresponding to $\sigma_{8,\rm IRAS} = 0.68 \pm 0.05$. The 1-$\sigma$ errors should be treated cautiously and are discussed in the paper.

We also relax the gravitational instability assumption, to find a more general determination of the velocity power spectrum required to reconcile the anisotropic redshift space map with the assumed isotropic real-space map.

## 1 INTRODUCTION

An accurate three-dimensional map of the galaxy distribution would be enormously valuable for cosmology, but the lack of an accurate estimator of distance precludes this. Despite best efforts, the errors in the Tully-Fisher and Faber-Jackson distance indicators mean that the best one can do is to use the recession velocity $v$, or redshift, of a galaxy, and to assign its distance $s$ by using the Hubble expansion law $v = H_0 s$, where $H_0$ is the Hubble parameter. The resulting 'redshift space' map is of course not perfect, because galaxies are not necessarily moving precisely with the idealised expansion of a homogeneous universe. Density inhomogeneities have associated with them peculiar velocities, which introduce a radial distortion between the true (real-space) map and the redshift space map. This distortion is, of course, inconvenient: for example, extracting the power spectrum of density fluctuations from a galaxy redshift survey becomes a non-trivial task. However, the distortion itself can be exploited, as the magnitude of the distortion is dependent on the density parameter of the Universe $\Omega_0$, if fluctuations grow by gravitational instability. Constraining $\Omega_0$ is one of the main aims of this paper.

In the most general terms, we can divide the methods for constraining $\Omega_0$ with peculiar velocities into two catagories which we shall label dynamical and statistical. In the former catagory perturbations in the density field are related to the peculiar velocity field through the continuity equation. Notable examples of this type are dipole studies (e.g. Lynden-Bell et al. 1989, Rowan-Robinson et al. 1990, Strauss et al. 1992), comparing the absolute motion of the local group to the gravitational field implied by redshift-surveys; the POTENT analysis (Bertschinger & Dekel 1989,

Dekel et al. 1993), making a point-by-point comparison of the reconstructed density field with redshift-maps; and the inverse problem of comparing the predicted to measured radial, relative peculiar velocities from redshift space maps (Kaiser et al. 1991). The analysis is complicated by the need to correct the density field for redshift-distortions, and in the latter two cases, the need to know what the radial peculiar velocities actually are. Given the uncertainties in redshift-independent distance indicators, and an ongoing debate over the effects of Malmquist bias, it is valuable to have a method which is independent of peculiar velocity measurements.

The need to measure peculiar velocities can be overcome within a statistical framework, and hence offers a powerful alternative to the dynamical methodology. The redshift-distortion introduces spurious terms into the statistical properties of the galaxy density field, which can be measured and used to constrain $\Omega_0$. In this paper we shall develop the statistical approach.

Coherent galaxy motion has the effect of locally transforming the real space coordinate system:

$$(r, \theta, \phi) \rightarrow (s, \theta, \phi) = (r, \theta, \phi) + (u, 0, 0), \tag{1}$$

where radial distances are estimated from Doppler redshifts. Here $u \equiv \mathbf{v}(\mathbf{r}).\hat{\mathbf{r}}/H_0$ is the radially projected galaxy peculiar displacement. An order of magnitude estimate suggests that the fractional change in the linear density field due to this transformation will be $\sim \beta \equiv \Omega_0^{0.6}/b$, the ratio of linear growth to galaxy bias parameter[*]. Hence, by measuring this change in the density field we can hope to constrain $\beta$.

In a seminal paper, Kaiser (1987; see also Lilje & Efstathiou 1989, McGill 1990) has given an approximate analysis of the distortion effect, which has been the basis for subsequent work. The Kaiser analysis consists of decomposing the density field into plane-waves, arriving at the linear expression

$$\delta^s(k_{ln}) = (1 + \beta \mu^2)\delta(k_{ln}), \tag{2}$$

in agreement with our approximate result, where $\mu$ is the cosine of the angle between the plane–wave and the line of sight. For high $\Omega_0$, as expected from inflation, and low galaxy bias, the projection into redshift space is non-negligible – the redshift space density differs from the the real-space density field by factors of order unity.

The distortion equation is derived assuming the gravitational instability hypothesis – now on a firm footing since the discovery of small initial fluctuations implied by the COBE–DMR observations, but can be generalised to models with non-gravitational velocities, provided that they are curl-free. However, the Kaiser equation can really only be applied universally if the survey subtends a small angle, so that the radial distortion can be approximated by a distortion along one axis. For a large-angle survey, one may be able implicitly to split the survey into small, independent volumes, but this will surely not be possible if the wavelength of interest subtends a large angle. Herein lies the problem of the Fourier approach: one wants to analyse short wavelengths, so the small-angle approximation may hold, but it is these short wavelengths which are difficult to analyse because they are nonlinear. With current survey depths, it is by no means obvious if there are any wavenumbers at all which are both linear and small-angle (cf Cole, Fisher & Weinberg 1994).

Further problems arise due to the choice of decomposition into Cartesian Fourier modes, a common choice for calculating the power spectrum from the galaxy distribution (we shall have more to say about statistics below). As the distortion is purely radial, we expect mixing of the Fourier modes (Zaroubi & Hoffman 1994). This will also arise as a result of finite and nonuniform sampling of the galaxy density field. Generally these effects can be reduced to independent radial and angular selection, the latter arising from incomplete sky-coverage. The redshift space distortion only interferes with the radial component, producing density distortions and surface effects. Finally, we would expect a purely radial distortion to destroy the statistical homogeneity and isotropy of the density field, whereas the Kaiser equation, while destroying isotropy, preserves homogeneity.

In this paper we re-examine the effect of redshift-distortion on a finite, inhomogeneously-sampled galaxy distribution using a decomposition into spherical Bessel functions and spherical harmonics, a formalism which was suggested for a somewhat different problem by Binney & Quinn (1991), and whose use for analysing galaxy catalogues was noted by Lahav (1993). The effects of redshift distortions in this description have been found independently by Fisher et al (1994c) and applied to reconstructing the real-space density field.

The method presented in this paper extends work in spherical harmonics in cosmology which goes back to Peebles (1973), and which includes work on analysis of near all-sky catalogues (Fabbri & Natale 1990, Scharf et al. 1992, Scharf & Lahav 1993) and velocity fields (Regős & Szalay 1989). The fact that redshift distortions affect the observed harmonics was noted by Scharf & Lahav (1993), and the distortion was exploited, without making a full decomposition of the field as here, to estimate $\beta$ by Fisher et al. (1994a).

Spherical coordinates are clearly the natural system to use, given the nature of the distortion, and a number of distinct advantages will become apparent. The two main advantages of the approach in this paper are firstly that the *only* approximation which one has explicitly to make is that the distortion is linear[†]. Since this method is essentially a power spectrum approach, it is straightforward to select only those modes which have long enough wavelength still to be in the linear regime. It therefore avoids the difficulties of correlation function methods (cf Hamilton 1992, Davis & Peebles 1983, Fisher et al. 1994b) in which the separation into linear and nonlinear regimes is not so clear-cut, and which usually involve noisy integrations of the correlation function out to large distances.

---

[*] Note that all methods which relate deviations from uniform density to deviations from uniform expansion are bound (at least within linear theory) to give answers for $\Omega_0$ which are uncertain because of ignorance of how the measured (number) overdensity of galaxies is related to the mass overdensity $\delta\rho$. The fractional number overdensity $\delta n/n$ is assumed in simple analyses to be $b\delta\rho/\rho$, where $b$ is a uniform bias parameter. In our analysis $b$ is defined in terms of the power spectrum. This is a weaker condition than relating number densities, as only the long wavelength modes used in our analysis are assumed to have a simple amplitude enhancement.

[†] A further approximation concerning discreteness of modes is implicit in our choice of boundary condition, similar to the usual practice of performing Fourier expansions in a cube of finite size.

The second major advantage is that this formalism offers a method of estimating $\beta$ and the *real-space* power spectrum itself. This facet is not pursued fully in this paper, but it is an important point. One of the major goals of cosmology in the next few years will be to measure the power spectrum on scales in the large gap between current measurements and those probed by microwave background experiments such as COBE. With reasonable depth surveys, one will have to consider the spherical nature of the distortion.

There are other minor advantages of this formalism: in a spherical coordinate system we can separate out the angular and radial effects and treat them independently, thus allowing us to deal with surface effects by imposing suitable boundary conditions. In addition we may calculate the change of phase information, as well as amplitude, of the harmonic coefficients. And finally, in applications where the distribution of objects sample cosmologically significant volumes, e.g. in the case of measuring structure in quasar surveys, curvature effects become important. In the framework of a spherical harmonic decomposition the radial component can be generalised to account for curvature effects, given a cosmological model.

In order to constrain cosmological parameters, we need to compare the anisotropic redshift space pattern of galaxy clustering with the expected underlying distribution. As we have no direct information as to the actual pattern, this must be done statistically. To proceed we need an *a priori* assumption for the underlying density distribution. The central limit theorem lends weight to the Random Gaussian Field Hypothesis, either from general considerations of the sum of many independently generated perturbations, or from inflationary theory. This also has observational support from the Rayleigh distributions of powers (Feldman et al. 1994), topology studies (Gott et al. 1986) and studies of skewness (Juskiewicz 1993). The distribution is then fully specified by the (isotropic) power spectrum of perturbations, $P(k \equiv |\mathbf{k}|)$. This can be either estimated along with $\beta$, from the data using a likelihood analysis, or put in by hand to find the maximal value of $\beta$ for a given model.

The form of the paper is set out as follows. In Section 3 and 4 we formulate the description of redshift space density perturbations in the language of spherical harmonics. This leads to a simple linear matrix relation between the modes of the underlying density field and the observed fluctuations in galaxy density in redshift–space. The stochastic properties of the observed galaxy density distribution are quantified using their covariance matrix in Section 4, including the shot noise contribution from finite sampling. We also consider here the effects of small scale, nonlinear velocity dispersion in redshift space. In Section 5 we expand on the stochastic properties of the redshift space density field by calculating the distribution of density fluctuations. Practical schemes for weighting the data are also considered in Section 5. The formalism is applied to a likelihood analysis of N-body and IRAS 1.2 Jy datasets in Section 6, using a parametric form for the underlying power spectrum. We also discuss in Section 6 how other observations, of the microwave background radiation, and of the abundances of rich clusters, may be combined with our analysis to constrain $\Omega_0$ itself.

### 1.1 Outline of the method

The method of analysis described in the paper consists of the following:

- Expansion of the redshift-space density field in spherical harmonics and spherical Bessel functions (equation 13).
- Conversion of real-space coefficients to redshift space, assuming linear theory, and taking into account incomplete sky coverage, selection function and real-to-redshift-space distortion (equation 18).
- Assuming an isotropic model power spectrum in real space, calculating the covariance matrix of the model coefficients in redshift space.
- Using the maximum likelihood technique to estimate the distortion parameter $\beta \equiv \Omega^{0.6}/b$ and the underlying real-space power spectrum.

## 2 PECULIAR VELOCITY FIELDS IN LINEAR THEORY

In a redshift space map, galaxies are placed at a position $\mathbf{s}=(s,\theta,\phi)$, where the distance coordinate $s$ is the recession velocity divided by the Hubble constant $H_0$. In general this is not the true distance because the galaxy may have a peculiar velocity $\mathbf{v}$. The redshift space position is then related to the real-space position $\mathbf{r}$ by

$$\mathbf{s}(\mathbf{r}) = \mathbf{r}\left[1 + \frac{(\mathbf{v} - \mathbf{v}_0) \cdot \mathbf{r}}{H_0 r^2}\right] \quad (3)$$

where $\mathbf{v}_0$ is the peculiar velocity locally.

In linear theory, the growing-mode of gravitational perturbations is curl-free (e.g. Peebles 1980), and the peculiar velocity may be expressed as the gradient of a velocity potential: $\mathbf{v} = -\nabla\Psi$. $\Psi$ is related to the mass overdensity $\delta_\rho \equiv \delta\rho/\rho$ by $\nabla^2\Psi = H_0 f(\Omega_0)\delta_\rho$ where $f(\Omega_0) \simeq \Omega_0^{0.6}$ is the logarithmic growth rate with respect to scale factor of the growing-mode perturbation.

## 3 SPHERICAL HARMONIC FORMALISM

### 3.1 Introduction and goals

In this section we set out the main equations relating the density field in redshift–space with the underlying density perturbations, and define the notation used throughout this paper. The goal is to relate the transform coefficients of a weighted redshift space survey to the transform of the actual density field in real space. The main result of this analysis is that the fields are related by a convolution, cast as a matrix relation between real and redshift space density modes. We begin by discussing the spherical harmonic formalism.

### 3.2 Spherical Harmonics

Spherical harmonics arise in physics as a result of their use as eigenfunctions of the Laplacian operator in systems with some spherical symmetry. Hence they can form a complete set of orthonormal basis functions. Following Binney

& Quinn (1991) we expand scalar quantities in spherical harmonics via the Bessel–Fourier relation

$$A(\mathbf{r}) = \sum_{lmn} c_{ln} A_{lmn} j_l(k_{ln} r) Y_{lm}(\theta, \phi), \quad (4)$$

$$A_{lmn} = c_{ln} \int A(\mathbf{r}) j_l(k_{ln} r) Y_{lm}^*(\theta, \phi) \, d^3 r, \quad (5)$$

where $c_{ln}$ is a normalisation factor, chosen to appear symmetrically in the transform and its inverse. $j_l(y)$ is the $l^{th}$ order spherical Bessel function, related to ordinary Bessel functions by $j_l(z) = \sqrt{\pi/(2z)} J_{l+\frac{1}{2}}(z)$, and $Y_{lm}(\hat{\mathbf{r}})$ is the Spherical Harmonic function of order $l$ and $m$, in the angular direction $\hat{\mathbf{r}}$. A number of conventions exist for spherical harmonics; we adopt that used by Binney & Tremaine (1987):

$$Y_{lm}(\theta, \phi) = \sqrt{\frac{2l+1}{4\pi} \frac{(l-|m|)!}{(l+|m|)!}} P_l^{|m|}(\cos\theta)$$
$$\times \exp(im\phi) \times \begin{cases} (-1)^m & m \geq 0 \\ 1 & m < 0 \end{cases} \quad (6)$$

The wavenumbers, $k_{ln}$, are chosen subject to a boundary condition on $A$ on the surface of a sphere of large radius $r_{max}$ (see Section 3.3). Note that both the $c_{ln}$ and $k_{ln}$ depend on $l$ as well as $n$.

In the notation of Dirac, we can represent the scalar product of the vector basis of configuration space with spherical harmonic space by (Binney & Quinn 1991)

$$\langle \mathbf{r} | lmn \rangle = c_{ln} k_{ln} j_l(k_{ln} r) Y_{lm}(\hat{\mathbf{r}}), \quad (7)$$

### 3.3 Boundary Conditions and Eigenvectors

As the potential field of the density distribution must satisfy Poisson's equation, unique solutions exist only for Dirichlet or Neumann conditions at the boundary (Jackson 1975). To first order we can avoid surface terms on the boundary of the survey by imposing Neumann boundary conditions, where the first derivative of the potential field normal to the boundary vanishes. Physically this requires that the velocity field be zero on the survey boundary, and so there is no distortion.

The surface condition of zero potential gradient implies that in spherical harmonic space the wavenumbers $k_{ln}$ are chosen such that

$$\frac{d}{dr} j_l(k_{ln} r) \bigg|_{r=r_{\max}} = 0, \qquad \forall \, l, n, \quad (8)$$

on the boundary. This is a somewhat different assumption from that used by Taylor and Rowan–Robinson (1993) when reconstructing the density field from redshift space, who set the potential to zero on and beyond the boundary. This ensured that the survey was treated as an isolated system. It also differs from the boundary condition used by Lahav (1993) and Fisher et al. (1994c), who required the potential to have a continuous derivative at the boundary. In our case, the selection function is very small at the boundary, so the results should not be sensitive to the choice of boundary condition. We have tested that the boundary conditions do not affect our results by placing the boundary at 200 and 250 $h^{-1}$ Mpc, and find consistent results.

For a finite sphere, we must introduce a normalisation constant to preserve normality. This can be calculated using the orthogonality properties of the harmonic basis (see Appendix C):

$$c_{ln} = \frac{2 k_{ln}^{3/2}}{\sqrt{\pi \{[1/4 + k_{ln}^2 r_{\text{rmax}}^2 - (l+1/2)^2] J_{l+1/2}^2(k_{ln} r_{\text{rmax}})\}}} \quad (9)$$

as the normalisation constant. A peculiarity arises with this choice of boundary condition; it can be shown (Fisher et al. 1994c) that the mean value of a field with the expansion (4) must be zero. This is dealt with by adding the mean value to the sum $A \to \langle A \rangle + \sum_{lmn}$. Since we shall be concerned with fluctuations around the mean, the extra term cancels.

### 3.4 Compact Notation

In later equations, the explicit reference to $lmn$ when referring to a particular mode is somewhat cumbersome. Hence, we shall on occasion use Greek indices as a shorthand for modes, keeping in mind the implicit relation $\mu \to (l, m, n)$. This is related to other representations of scalar quantities by

$$A_{lmn} = \langle A | lmn \rangle = A_\mu. \quad (10)$$

## 4  REDSHIFT SPACE DISTORTIONS

### 4.1 Transformations to redshift space for inhomogeneous samples

Having laid out the basic transformations, we shall now turn our attention to the relationship between the transform of the *observed* density field, and the true density field. The former is modified in two ways: firstly, because of the selection function, the mean density of galaxies drops off with distance; secondly, the observed distances are redshift distances, not real distances. A further complication arises if we wish to weight the galaxies depending on their redshift (in some applications, inverse selection function weighting may be appropriate, e.g. in reconstructions of the real-space density field, Fisher et al. 1994c). We will identify the observed coefficients by $\tilde{\rho}_{lmn}$, and the coefficients of the underlying real density field by $\rho_{lmn}$.

In general the observed redshift density field can be defined as a sum of Dirac delta functions,

$$\rho(\mathbf{s}) = \sum_i \delta_D(\mathbf{s}(\mathbf{r}, \mathbf{v}) - \mathbf{r}_i) \delta_D(\mathbf{v} - \mathbf{v}_i), \quad (11)$$

projected from the phase space distribution of galaxies. This can also be expressed in terms of fluctuations about a background mean density field, which will in general be spatially varying due to selection criteria such as flux limits and unobserved regions of sky. We shall denote this generalised background by $\rho_0(\mathbf{r})$. Hence the observable density field is now expressible as

$$\rho(\mathbf{s}) = \rho_0(\mathbf{r})(1 + \delta(\mathbf{s})). \quad (12)$$

The background density, $\rho_0$, may in fact be a function of the real and redshift–space coordinate systems, as the condition for inclusion into a catalogue may be functions of either system. Here we shall only consider real-space selection criteria.

Applying the harmonic transform to the redshift density field, we construct

$$\tilde{\rho}_{lmn} = c_{ln} \int w(s)\rho(\mathbf{s})j_l(k_{ln}s)Y_{lm}(\theta,\phi)d^3s \tag{13}$$

where we have included an optional weighting function $w(s)$. Using the continuity condition $d^3s\,\rho(\mathbf{s}) = d^3r\,\rho(\mathbf{r})$, we can transform the integral to real space;

$$\tilde{\rho}_{lmn} = c_{ln} \int w(s)\rho(\mathbf{r})j_l(k_{ln}s)Y_{lm}(\theta,\phi)d^3r. \tag{14}$$

where the distortion is now restricted to the argument of $j_l(k_{ln}s)$ and the weighting function $w(s)$. We see also that our choice of boundary condition has avoided the introduction of surface terms into this expression. In general, we should expect that these would be negligible in any case, being dominated by the effects of galaxy selection near the survey boundary.

While expression (14) is exact it does not represent a complete solution to our problem, as the relationship between **s** and **r** has not been supplied. To complete the analysis we shall proceed by perturbation expansion of $j_l(k_{ln}s)$ to first order. From Section 2 we see that the velocity field can be related to the gravitational potential by a gradient operator, and that the redshift displacement is the radial projection of this. Using our results (Appendix A) for the vector spherical harmonics we see that

$$
\begin{aligned}
u(\mathbf{r}) &= -\hat{\mathbf{r}} \cdot \nabla \left( \frac{\Omega_0^{0.6}}{\nabla^2} \delta_\rho(\mathbf{r}) \right), \\
&= \Omega_0^{0.6} \sum_{lmn} c_{ln} k_{ln}^{-1} \delta_{lmn} j_l'(k_{ln}r) Y_{lm}(\theta,\phi),
\end{aligned}
\tag{15}
$$

is the radial potential field term of the velocity field and where $j_l'(z) \equiv dj_l/dz$. $\delta_{lmn}$ is the transform of the fractional galaxy number overdensity field $\delta(\mathbf{r})$, and we shall assume that the galaxies are biased tracers of the density field, in the sense that $\delta_{lmn} = b\,(\delta_{lmn})_\rho$. This condition is only assumed to apply for the modes used in the analysis. In particular, no assumptions need to be made about the short-wavelength, highly nonlinear modes, as these will not significantly affect the long-wavelength modes analysed here. With this assumption, $\Omega_0^{0.6} \to \beta \equiv \Omega_0^{0.6}/b$, if we assume that $b$ is independent of scale $k$. For $l > 0$, $\delta_{lmn}$ is simply related to the transform of the density field $\rho_{lmn}$ by $\rho_{lmn} = \bar{\rho}\delta_{lmn}$, where $\bar{\rho}$ is the mean number density[‡].

Note that the modes are independent in real space:

$$\langle \delta_{lmn} \delta_{l'm'n'}^* \rangle = b^2 P(k_{ln}) \delta_{ll'}^{\rm K} \delta_{mm'}^{\rm K} \delta_{nn'}^{\rm K} \tag{16}$$

where $P(k)$ is the power spectrum of mass fluctuations, and $\delta^{\rm K}$ is the Kronecker delta function.

Returning to the expansion of the observed redshift space density field, we can expand $j_l(k_{ln}s)$ to first order by

$$
j_l(k_{ln}s) \simeq j_l(k_{ln}r) + u(\mathbf{r})\frac{d}{dr}j_l(k_{ln}r),
$$

---

[‡] Interestingly, we do not need to assume gravitational instability; we simply need to determine the velocity field for which the real-space density power spectrum is isotropic. This is discussed in Section 6.

$$
\begin{aligned}
&\simeq j_l(k_{ln}r) + \\
&\quad \beta \sum_{l'm'n'} c_{l'n'} \frac{\delta_{l'm'n'}}{k_{l'n'}} j_{l'}'(k_{l'n'}r) j_l'(k_{ln}r) Y_{l'm'}.
\end{aligned}
\tag{17}
$$

Consider now the effects of a inhomogeneous background density. Expanding out the scalar perturbations, $\delta(\mathbf{r})$, and using the orthonormality of $j_l(k_{ln}r)$ and $Y_{lm}(\theta,\phi)$ (see Appendix C), we find that

$$\tilde{\rho}_{lmn} = (\rho_0)_{lmn} + \sum_{l'm'n'}(\Phi_{ll'nn'}^{mm'} + \beta V_{ll'nn'}^{mm'})\delta_{l'm'n'}, \tag{18}$$

where

$$(\rho_0)_{lmn} = c_{ln} \int \rho_0(r)w(r)j_l(k_{ln}r)Y_{lm}(\theta,\phi)d^3r \tag{19}$$

is the transform of the observational, weighted mean density field. In the general expression for $\tilde{\rho}_{lmn}$ the transition matrices are defined as

$$
\begin{aligned}
\Phi_{ll'nn'}^{mm'} &\equiv \langle lmn|\rho_0 w|l'm'n'\rangle, \\
&= c_{ln}c_{l'n'} \int \rho_0(\mathbf{r})w(r) \times \\
&\qquad j_l(k_{ln}r)Y_{lm}(\theta,\phi)j_{l'}(k_{l'n'}r)Y_{l'm'}^*(\theta,\phi)d^3r,
\end{aligned}
\tag{20}
$$

and

$$
\begin{aligned}
V_{ll'nn'}^{mm'} &\equiv \frac{c_{ln}c_{l'n'}}{k_{ln}^2} \int \rho_0(\mathbf{r})\frac{d}{dr}[w(r)j_l(k_{ln}r)] \\
&\qquad \times Y_{lm}(\theta,\phi)j_{l'}'(k_{l'n'}r)Y_{l'm'}^*(\theta,\phi)d^3r.
\end{aligned}
\tag{21}
$$

In a more compact notation, we arrive at a final expression relating the undisturbed density field with the observable redshift distorted density perturbations, $D_\mu \equiv (\tilde{\rho}_{lmn} - (\tilde{\rho}_0)_{lmn})/\bar{\rho}$;

$$D_\mu = \sum_\nu \Lambda_\mu^\nu \delta_\nu, \tag{22}$$

where

$$\Lambda_\mu^\nu \equiv \Phi_\mu^\nu + \beta V_\mu^\nu. \tag{23}$$

Equation (22), in this its most general form, is one of the main results of this paper. In comparison with the Kaiser equation, we note that redshift distortions causes the mixing of modes, as we expected to happen from the general grounds discussed in the introduction. This expression is exact to first order, and correctly deals with the finite extent of the redshift survey.

### 4.2 Division into Radial and Angular Selection

One of the main advantages of using spherical coordinates is the natural division into radially dependent and angularly dependent functions. In application to large angle redshift surveys this is particularly useful. Thus we can decompose $\rho_0(\mathbf{r})$ into

$$\rho_0(\mathbf{r}) = W(\theta,\phi)\rho_0(r), \tag{24}$$

where $W(\theta,\phi)$ is an angular selection function, and $\rho_0(r)$ now embodies the radial selection function.

In the case of all–sky redshift surveys such as the IRAS QDOT and Berkeley 2-Jy and 1.2-Jy surveys, there is a region around the galactic plane, the zone of avoidance, that is excluded from the survey due to source confusion. This masking introduces, for example, a large quadrupole term in to the distribution of sources, which is then mixed with other modes by the redshift–space distortion. With this separable form for the selection function, the transition matrices become (cf Peebles 1973, Scharf et al. 1992)

$$\begin{aligned} \Phi_{ll'nn'}^{mm'} &= W_{ll'}^{mm'} \Phi_{ll'}^{nn'} \\ V_{ll'nn'}^{mm'} &= W_{ll'}^{mm'} V_{ll'}^{nn'} \end{aligned} \quad (25)$$

(the summation convention is suspended) and

$$\begin{aligned} \Phi_{ll'}^{nn'} &\equiv c_{ln}c_{l'n'}\int \rho_0(r)w(r)j_{l'}(k_{l'n'}r)j_l(k_{ln}r)r^2 dr \\ V_{nn'}^{ll'} &\equiv \frac{c_{ln}c_{l'n'}}{k_{l'n'}^2}\int \rho_0(r)\frac{d}{dr}[w(r)j_{l'}(k_{l'n'}r)]j_l'(k_{ln}r)r^2 dr \\ W_{ll'}^{mm'} &\equiv \int_\Omega Y_{l'm'} M(\Omega) Y_{lm}^* d\Omega, \quad \Omega \equiv (\theta,\phi). \end{aligned} \quad (26)$$

For a complete all-sky survey, the expressions simplify, since $W$ reduces to delta-functions: $W_{ll'}^{mm'} = \delta_{ll'}^{\rm K}\delta_{mm'}^{\rm K}$. So, in the absence of a mask, the mixing is between modes with the same $l$ and $m$, with considerable simplification.

### 4.3 The Covariance Matrix of the $D_{lmn}$'s

In the general case, we can compare theoretical models with the observations by constructing ensemble averages;

$$\langle D_\mu D_\nu \rangle = \frac{1}{2}\sum_\alpha (\Phi_\mu^\alpha + \beta V_\mu^\alpha)(\Phi_\nu^\alpha + \beta V_\nu^\alpha) P(k_\alpha) + \Lambda_{\mu\nu}^0 \quad (27)$$

where, in practice, we split the coefficients into real and imaginary parts, which are independent provided that the mask mixing matrices $W$ are real (this is true for a zone of avoidance mask, and very nearly true for the IRAS mask). This explains the factor $\frac{1}{2}$ in the power. The last term is the shot noise, which can be obtained by the methods of Peebles (1973);

$$\begin{aligned} \Lambda_{\mu\nu}^0 &= c_{ln}c_{l'n'}\int \rho_0(r)w^2(r)j_l(k_{ln}r)j_{l'}(k_{l'n'}r)r^2 dr \times \\ & \int \mathcal{P}_\mu(Y_{lm}(\Omega))M(\Omega)\mathcal{P}_\nu(Y_{l'm'}^*(\Omega))d\Omega. \end{aligned} \quad (28)$$

Here $\mathcal{P}_{\mu,\nu}$ represent real or imaginary parts, depending on whether $D_{\mu,\nu}$ are real or imaginary. In order to calculate the shot noise contribution, we use the fact that Tesseral harmonics (spherical harmonics of the type $\mathcal{R}e$ or $\mathcal{I}m[Y_l^m(\Omega)]$) may be written, for example, $\mathcal{R}e[Y_l^m(\Omega)] = \frac{1}{2}[Y_{lm}(\Omega) + Y_{lm}^*(\Omega)]$, which allows us to calculate the angular integral in terms of four $W_{ll'}^{mm'}$ matrices.

Equation (27) will form the basis of estimation of parameters via likelihood methods, but it is instructive to pause and consider it for a moment. Essentially we make the comparison in the observational domain, assuming a power spectrum $P(k)$ and a value of $\beta$. Notice that the observed quantities depend on both of these, so there is hope that a maximum likelihood method would yield information on the power spectrum and on $\beta$. Indeed, for certain weighting functions, the mixing matrices are not very broad, so estimation of the power spectrum may be possible without great uncertainty. This will be the subject of a subsequent paper; here we assume a parameterised form for the power spectrum.

### 4.4 Small–Scale Velocity Structure

The small–scale velocity field of galaxies introduces an extra fine structure in redshift space. Once overdensities collapse and subsequently virialize, their velocity distribution relaxes towards a Maxwellian distribution. Galaxies in these structures are scattered radially about their true positions in the cluster producing extended structures in redshift–space. These have been called "Fingers of God" due to their radial alignment. This random behaviour is strongly nonlinear in origin, but works in the opposite sense to the linear compression of large–scale features. Small–scale structures are smeared out, resulting in the strong aliasing of power from the higher to smaller, linear wavenumbers.

We can mimic the effect of the small-scale velocity structure in redshift–space by randomly assigning each galaxy an additional peculiar velocity drawn from a Maxwellian distribution (cf Peacock & Dodds 1994), and projecting radially. This is approximate, since it assumes that the velocity dispersion is independent of location. Formally the effect can be calculated by first introducing a small, random perturbation into the position of each galaxy, and averaging:

$$D'(\mathbf{r}) = \langle \rho_0(\mathbf{r})\delta[\mathbf{s}'(\mathbf{r})]\rangle_\varepsilon, \quad (29)$$

where the redshift–space coordinates are displaced by

$$\mathbf{s}(\mathbf{r}) \longrightarrow \mathbf{s}'(\mathbf{r}) = \mathbf{s}(\mathbf{r}) + \varepsilon(\mathbf{r}). \quad (30)$$

Here $\varepsilon(\mathbf{r})$ is a Maxwellian distributed random variable, with $\langle \varepsilon(\mathbf{r})\rangle = 0$ and $\langle \varepsilon_i(\mathbf{r}_1)\varepsilon_j(\mathbf{r}_2)\rangle = \sigma^2\delta_{ij}\delta_D(\mathbf{r}_1 - \mathbf{r}_2)$, where $\sigma H$ is the assumed uncorrelated 3-D velocity dispersion.

It is straightforward to show that the modes of the perturbed observable galaxy density are related to the unperturbed field in redshift–space by a convolution. We shall leave the outline of this to Appendix B, but quote the result, which is

$$D'_{lmn} = S_{ll'nn'}^{mm'} D_{l'm'n'}, \quad (31)$$

where the cluster scattering matrix is given by

$$\begin{aligned} S_{ll'nn'}^{mm'} &\equiv \frac{c_{ln}c_{l'n'}}{V^2}\iint p(\mathbf{r} - \mathbf{y})j_l(k_{ln}r)Y_{lm}(\theta_r,\phi_r) \\ & \times j_{l'}(k_{l'n'}y)Y_{l'm'}(\theta_y,\phi_y)r^2 dr\, d\Omega_r\, y^2 dy\, d\Omega_y\,. \end{aligned} \quad (32)$$

To find the scattering matrix suitable for projections along the radial direction, we decompose the Maxwellian velocity distribution, $p(\mathbf{r} - \mathbf{y})$, into a radial and angular dependent part;

$$\begin{aligned} p(\mathbf{r} - \mathbf{y}) &= \frac{e^{-(\mathbf{r}-\mathbf{y})^2/(2\sigma^2)}}{(2\pi)^{3/2}\sigma^3} \\ &= \left[\frac{e^{-(r-y)^2/(2\sigma^2)}}{\sqrt{2\pi}\sigma}\right]\left[\frac{e^{-ry(1-\mu)/\sigma^2}}{2\pi\sigma^2/ry}\right]\left[\frac{1}{ry}\right], \end{aligned} \quad (33)$$

where we identify the angular dependent term in the expansion as a negative exponential distribution. Allowing the width of this term to vary as $\sigma \to \sigma'$, we can collapse this distribution into a pencil beam. As $\sigma' \to 0$, we find

$$\frac{e^{-ry(1-\mu)/\sigma'^2}}{2\pi\sigma'^2/ry} \xrightarrow{\sigma' \to 0} \delta_D(1-\mu)/(2\pi). \qquad (34)$$

Applying this constraint to the scattering matrix equation (33), we arrive at the symmetric matrix for a radially projected random perturbation in galaxy positions;

$$\begin{aligned} S_{ll'nn'}^{mm'} &= \frac{c_{ln}c_{l'n'}\delta_{ll'}^K \delta_{mm'}^K}{V\pi} \int\int \\ &\times \frac{e^{-(r-y)^2/(2\sigma^2)}}{\sqrt{2\pi}\sigma} j_l(k_{ln}r) j_{l'}(k_{l'n'}y) r\, dr\, y\, dy. \end{aligned} \qquad (35)$$

Note that we absorb the mask into the translation from real to (unsmoothed) redshift space, so the smoothing is a delta function in $m$ and $m'$. This random scattering of the observed density field can be folded into our calculations for the real underlying modes of the density field by the final expression

$$D'_{lmn} = S_{ll''nn''}^{mm''} W_{l''l'}^{mm'} (\Phi_{l''l'}^{n''n'} + \beta V_{l''l'}^{n''n'}) \delta_{l'm'n'}, \qquad (36)$$

where summation is implied.

Having specialised to the case of radial random fluctuations in the positions of galaxies, it apparent that the effects of measurement errors in galaxy redshifts can also be accounted for by adding this to the velocity variance in quadrature.

### 4.5 Low-order modes and uncertainties in the Local Group Velocity

The relationship between the overdensity coefficients $\delta_{lmn}$ and the density coefficients is simply $\rho_{lmn} = \bar{\rho}\delta_{lmn}$, except for the $l = 0$ modes, for which the radial selection function is involved (see Fisher et al. 1994c for more discussion of this point). In addition, if we change the frame in which we measure the peculiar velocities (perhaps by choosing the Local Group frame, and revising the Local Group velocity with respect to the microwave background), the coefficients remain unchanged, except for the dipole term. For these reasons, it is sensible to restrict one's attention to the quadrupole term and higher. If the survey is not all-sky, changing frames does affect $l \neq 1$ modes, since there is some mixing between modes, but these are small if the masked regions of the survey are small.

## 5 STOCHASTIC PROPERTIES

Having set out the formal relationship between individual harmonic modes of the underlying density field in real-space, and that of a set of modes measured in an observationally-limited redshift survey, we now move on to discuss the statistical properties of the observable density modes. Our main result here is that the observable density field no longer retains all the statistical properties of the true density field. In particular the usual assumption of statistical homogeneity and isotropy of the density field break down for a finite redshift survey. It is this fact that allows us to use a statistical approach to determine the amplitude of the distortion, and so measure $\beta$.

In Section 4.3, we calculated the covariance matrix of perturbations in redshift space, and related it to the power spectrum of the true density field. Using this relation, we may make a random phase hypothesis regarding the phases of the true density perturbations. This leads us to regard the harmonic modes of the true density field as Gaussian distributed. A fundamental property of Gaussian random fields is that the linear transform of the density field still obeys the Gaussian form. In Section 6, we shall use this property to measure the degree of distortion from statistical isotropy and homogeneity present in each mode of the observable density field.

### 5.1 Statistics of Observable density fields

The lowest order non-trivial statistic of the perturbations is the covariance matrix of each mode of the density field, $\langle D_{lmn} D_{l'm'n'} \rangle$, where angled brackets, $\langle \rangle$, denote ensemble averaging, and is given by equation (27). The last term in that equation is due to discrete shot noise. Clearly, as we shall assume that this is a random sampling of the density field by galaxies, this term is independent of redshift space distortion. However it is still affected by the finite volume and both angular and radial selection functions.

#### 5.1.1 *Recovery of Kaiser's result for $k \to 0$*

It is interesting to consider the relationship between the average redshift space power $P^s(k)$ and the real-space power $P(k)$, and to compare it with the analogous result of Kaiser (1987) for the distant-observer approximation ($P^s(k) = [1 + 2\beta/3 + \beta^2/5] P(k)$). Since the redshift distortion introduces mixing between modes, the ratio of $P^s(k)$ to $P(k)$ depends on the power spectrum, in contrast to Kaiser's analysis:

$$P^s(k) = \langle |D_\mu|^2 \rangle = \left\langle \sum_\alpha (\Phi_\mu^\alpha + \beta V_\mu^\alpha)^2 P(k_\alpha) \right\rangle, \qquad (37)$$

where the averaging is done over bins in wavenumber, and we ignore shot noise, incomplete sky coverage and incoherent velocity dispersion. Multiplying out the square, and performing the summation, for a flat power spectrum, ($P(k) =$ constant), we calculate the coefficients of $\beta$ and $\beta^2$. These are plotted in Figure 1, along with the values 2/3 and 1/5 from Kaiser's analysis.

Two effects on the values of the coefficients that are not seen in the Kaiser terms become apparent. Firstly, it is apparent that edge effects at low $k_{ln}$ attenuate the amplitude of the modes. As we only have a finite sample, this expresses the fact that information about modes larger than the survey are not directly measurable.

The second effect is due to the small–angle approximation used to derive the Kaiser equation. This has the effect of pushing the range of validity of the Kaiser equation to higher $k_{ln}$ and $l$ modes. This is the same as saying that a mode in the linear regime subtending a small angle will have to be at a correspondingly larger distance. Both of these effect conspire to push the range of validity of the Kaiser equation to smaller scales, where we expect nonlinear coupling in the growth of structure to become dominant. If this is the case, then the range of applicability of the Kaiser equation may be somewhat limited.

## 5.2 Likelihood Analysis of the Observable density field

A natural condition for density perturbations formed in the early universe to satisfy is that of random phases. The Central Limit Theorem implies that these density fluctuations will be randomly Gaussian distributed, with all the necessary statistical information contained in the covariance, or correlation, function. In the case of a continuum density field, which has an infinite number of degrees of freedom, the distribution function is dependent on the values of the density field at all points, and is called a distribution functional.

The distribution functional for density perturbations expresses the probability for a given configuration of fluctuations to occur. This configuration can be quantified either in terms of the field amplitude at each point, or more compactly, in terms of the amplitude of each harmonic mode. Generalizing the case of a multivariate Gaussian, the probability distribution functional of the density field can be expressed as

$$P[\delta] = (\det A)^{-1/2} \exp\left(-\frac{1}{2}\int \delta(r) A^{-1}(r,y) \delta(y)\, d^3r d^3y\right), \tag{38}$$

where

$$(\det A)^{1/2} = \int D[\delta]\exp\left(-\frac{1}{2}\int \delta(r) A^{-1}(r,y) \delta(y)\, d^3r d^3y\right), \tag{39}$$

is the natural extension of the determinant to an infinite-dimensional function space. The Wiener measure, $D[\delta(\mathbf{x})]$, is defined in the limit as $\lim_{N\to\infty} \prod_{n=1}^{N} d\delta(\mathbf{x}_n)/(2\pi)^{n/2}$. The function $A(r,y) = \xi(|r-y|)$ is the correlation function of the density perturbations. An equivalent expression can be found by expanding the density field in terms of its harmonics, leading to the expression

$$P[\delta] = (\det A)^{-1/2} \exp\left(-\frac{1}{2}\sum_{lmn}\frac{\delta_{lmn}\delta^*_{lmn}}{P(k_{ln})}\right), \tag{40}$$

where we have used the statistical independence of the harmonic modes to simplify the summation.

The amplitudes $|\delta_{lmn}|$ follow a Rayleigh distribution, and the phases are random, or equivalently, the real and imaginary parts of $\delta_{lmn}$ are independently Gaussian distributed. We can use this to property to split the probability distribution functional of the density field into two independent parts, one for real terms and one for the imaginary terms;

$$\begin{aligned}
P[\delta] &= P[\mathcal{R}e(\delta_{lmn})]\, P[\mathcal{I}m(\delta_{lmn})], \\
P[\mathcal{R}e(\delta_{lmn})] &= (\det A_{\mathcal{R}e})^{-1/2}\exp\left(-\sum_{lmn}\frac{\mathcal{R}e(\delta_{lmn})^2}{2P(k_{ln})}\right), \\
P[\mathcal{I}m(\delta_{lmn})] &= (\det A_{\mathcal{I}m})^{-1/2}\exp\left(-\sum_{lmn}\frac{\mathcal{I}m(\delta_{lmn})^2}{2P(k_{ln})}\right),
\end{aligned} \tag{41}$$

where $A_{\mathcal{R}e}$ and $A_{\mathcal{I}m}$ are the correlation matrices of the real and imaginary coefficients respectively.

A general property of Gaussian random fields is that linear operations on the random variable are also Gaussian distributed. Our transformation equation (22) from the real density to the observed density field is of a linear form, and so the real and imaginary parts of $D_{lmn}$ are also Gaussian distributed. The linearity of equation (22) is essential here, and arises as a results of choosing spherical harmonics as our basis for expanding perturbations. In other bases, both the redshift projection and window function may be nonlinear operations, and our assumption of Gaussianity would be doubtful.

With this in mind, we are now in a position to construct a likelihood functional for the observed density field. Applying the transform (22) to the underlying density field, and using equation (27) for the correlation matrix of observable density modes, we find

$$\begin{aligned}
\mathcal{L}[D|\beta, P(k_{ln})] &= (\det C_{\mathcal{R}e})^{-1/2}(\det C_{\mathcal{I}m})^{-1/2} \\
&\times \exp\left(-\frac{1}{2}\sum_{lmn}\mathcal{R}e(D_{lmn})(C^{mm'}_{ll'nn'\mathcal{R}e})^{-1}\mathcal{R}e(D_{l'm'n'})\right) \\
&\times \exp\left(-\frac{1}{2}\sum_{lmn}\mathcal{I}m(D_{lmn})(C^{mm'}_{ll'nn'\mathcal{I}m})^{-1}\mathcal{I}m(D_{l'm'n'})\right).
\end{aligned} \tag{42}$$

as obtained by Scharf & Lahav 1993 and Fisher et al. 1994a. The inhomogeneity of the observed density field is manifest in the likelihood by the correlations between modes, which also contain valuable phase information.

In principle we can proceed in a number of ways; assume some functional form for the true power spectra, and maximize with respect to $\beta$, or allow some freedom in the shape of the spectrum and maximize $\mathcal{L}$ for a fixed $\beta$, or allow both the shape and $\beta$ to be determined simultaneously. In the next section we discuss and apply some of these possibilities.

It is worth noting at this point that the method we have outlined is in essence a power spectrum analysis. Aside from allowing us to represent compactly the amplitude and phase information of the density field in terms of fundamental modes, it allows us to separate the density perturbations into linear and nonlinear regimes. This is an essential feature, as the method outlined is exact only to first order. Even the inclusion of a correction term from the random velocities found in clusters only serves to suppress the contamination of small scale modes aliasing as linear modes. This highlights the general problem of nonlinear modes coupling with linear modes, due to selection functions mixed with redshift distortions. With careful choice of weighting functions for the data, the mixing can be made minimal. This contrasts strongly with methods based on the extraction of redshift space distortion from galaxy correlations, where the distinction between linear and nonlinear regimes is ambiguous.

However, a general feature of the statistical approach is the degeneracy that will arise in the parameter space of redshift distortion, $\beta$, and the intrinsic amplitude of density distortions. An increase in the distortion has the same effect on radial modes as that of increasing the amplitude of radial perturbations. However, homogeneity demands that an increase in power is distributed equally between all modes, regardless of phase. Hence the likelihood functional, (42),

### 5.3 Optimal Weighting of the data

In Section 4.2 we allowed ourselves the option of an arbitrary weighting scheme for each galaxy in the sample. This is a desirable feature, as is it not apparent that each galaxy in a sample subject to selection and bias contains an equal amount of information about the underlying density field. Indeed, it turns out to be essential to adopt a suitable weighting scheme, to avoid the convolutions arising from the selection function introducing coupling between too many different modes, which becomes computationally difficult to handle.

Given that we do have prior information about how the sample was selected, from the magnitude limit, window function, etc, it is reasonable that we use this information to optimally weight each galaxy when measuring a specific statistic.

A weighting scheme can be called optimal if it minimises the variance in the measured statistic, in which case it is called a minimum variance weighting scheme. Such a minimum in the variance of a statistic can be found by solving the functional equation

$$\frac{\delta}{\delta w(y)} \text{Var}[f(r)] = 0, \tag{43}$$

where $f$ is the statistic being measured. The variance of $f(r)$ can be calculated directly from the likelihood function using the approximate relation

$$\text{Var}[f(r)] \simeq -\left\langle \frac{\delta^2 \ln \mathcal{L}[f(y)]}{\delta f(r)^2} \right\rangle^{-1}, \tag{44}$$

where $-\ln \mathcal{L}$ is the error functional and $-1/\text{Var}[f]$ is called the amount of information. Applying this expression to the Gaussian random field model, we find

$$\text{Var}[f(r)] \simeq 2 \left(\frac{C}{C'}\right)^2, \tag{45}$$

where $C$ is the covariance matrix and $C' \equiv \delta C/\delta f$ is the functional derivative of $C$ with respect to $f(r)$.

For the moment we shall ignore the effects of redshift–space distortion, and concentrate on finding the minimum variance weighting function for a function that is distortion independent. After this we shall be in a better position to generalize the problem to include distortions.

Taking the diagonal elements of the covariance matrix only, $C_{ii} = \text{diag}\, C_{ij}$, and assuming that the power spectrum is approximately constant over the range of interest, we find that

$$C_{ln} \approx \int w^2(r)\phi^2(r) \left(\frac{1}{\phi(r)} + P(k_{ln})\right) j_l^2(k_{ln}r)\, d^3r, \tag{46}$$

where we have used the closure relation for spherical Bessel functions (Appendix C) to reduce the summation over modes.

The two statistics we are most interested in obtaining from redshift space are the power spectrum of density fluctuations, $P(k_{ln})$, and the distortion parameter, $\beta$. In the limit of no distortions, only the power spectrum is relevant. The variance of the power spectrum, measured over some small range in harmonics, is

$$\frac{\text{Var}\, P(k_{ln})}{P^2(k_{ln})} \approx$$

$$\frac{2\int w^4(r)\phi^4(r) \left|1 + \frac{1}{\phi(r)P(k_{ln})}\right|^2 j_l^4(k_{ln}r)\, d^3r}{|\int w^2(r)\phi^2(r)j_l^2(k_{ln}r)\, d^3r|^2}. \tag{47}$$

Taking the functional derivative with respect to the weighting function, $w(y)$, and remembering the functional relation $\delta w(\mathbf{x})/\delta w(\mathbf{y}) = \delta_D(\mathbf{x} - \mathbf{y})$, the minimum variance condition is

$$\frac{\delta}{\delta w} \ln \frac{\text{Var}\, P}{P^2} \approx$$

$$\frac{4\int w^3(r)\phi^4(r) \left|1 + \frac{1}{\phi(r)P(k_{ln})}\right|^2 j_l^4(k_{ln}r)\, d^3r}{\int w^4(r)\phi^4(r) \left|1 + \frac{1}{\phi(r)P(k_{ln})}\right|^2 j_l^4(k_{ln}r)\, d^3r}$$

$$+ \frac{4\int w(r)\phi^2(r)j_l^2(k_{ln}r)\, d^3r}{\int w^2(r)\phi^2(r)j_l^2(k_{ln}r)\, d^3r} = 0, \tag{48}$$

with solution

$$w(r) \approx \frac{1}{1 + P(k_{ln})\phi(r)}, \tag{49}$$

where we have ignored an unimportant amplitude. This coincides with the optimal weighting function derived somewhat differently by Feldman, Kaiser & Peacock (1993), although both derivations make use of the properties of the Gaussianity of the density field. This weighting scheme has the expected form of signal/(signal + noise) that one would expect from discrete data that can be decomposed into a Gaussian distributed signal and Poissonian shot noise. A big advantage of this weighting function is that the mixing matrices $\Phi_{ll'nn'}^{mm'}$ are much narrower in k-space than those for equal weighting, and fewer modes need to be considered (see figure 2). This form leads us to expect that a minimum variance weighting scheme can be found by applying the approximation

$$\frac{\text{Var}[f(r)]}{f^2(r)} \approx \left|\frac{d\ln C}{d\ln f}\right|^2, \tag{50}$$

where $C$ is approximated by equation (46). This allows us to extend the analysis to the case where $\beta \neq 0$. The problem is now much more difficult, since the $\beta$ term involves derivatives of the Bessel functions. In order to proceed, we assume that these derivatives obey an approximate orthogonality relation; since this is not quite true, the weighting functions are not quite optimal, but may still be useful. We assume further that $w(r)$ is locally a smooth power law with slope $\gamma$: $w(r) \simeq (r/r_0)^\gamma$. We can then repeat the derivation of of the weighting function for the power spectrum, to find

$$w_P(r) \approx \frac{(1 + \beta + \gamma\beta r_0/r)}{1 + (1 + \beta + \gamma\beta r_0/r)^2 P(k_{ln})\phi(r)}, \tag{51}$$

which in the special case of $\beta = 0$ reduces to equation (49). Applying the same formalism to the variance on $\beta$ yields

$$w_\beta(r) \approx \frac{\sqrt{(1 + \beta + \gamma\beta r_0/r)(1 + \gamma r_0/r)}}{1 + (1 + \beta + \gamma\beta r_0/r)^2 P(k_{ln})\phi(r)}. \tag{52}$$

In Figure 3(a) we plot the selection functions for QDOT,

the 1.2-Jy and 2-Jy IRAS surveys. Assuming that the power spectrum of perturbations can be approximated by a mean value, $\bar{P}(k)$, we show in Figure 3(b) the weighting functions corresponding to the 1.2-Jy catalogue. The solid curve shows uniform weighting, the Feldman, Kaiser, Peacock weighting scheme is shown by a dashed curve, while dot-dashed and dotted curves show the redshift–space corrected weighting schemes. All curves are normalized to unity at $r = 3 \times 10^4$ km$^{-1}$s.

Interestingly, the optimal weighting schemes for $P(k)$ and $\beta$ are very similar, and significantly different from the uncorrected weights. Two effects are can be seem. At low redshift, the $\gamma\beta r_0/r$ terms are dominant, and correct for the misplacing of the weights in redshift space. Beyond $r_0$, the dominant effect is the amplification of the power by a factor $1 + \beta$.

It is worthwhile clarifying some of the approximations used in deriving the two weighting functions, $w_P(r)$, and $w_\beta(r)$, and what they actually correct for. Our main assumption is that the minimum variance weights will be approximately of the form $\sim (C'/C)^2$, which is true for Gaussian distributed variables. Also we have assumed orthogonality between the derivatives of the spherical Bessel functions in order to derive equations (51) and (52). While this allows us to introduce factors of $1 + \beta + \gamma\beta r_0/r$, we have ignored the angular dependence of the distortion. However, as our weighting scheme is arbitrary, we are free to weight galaxies as we please, and so are only concerned with finding an approximately optimal weighting scheme. Its value can then be assessed by its influence on the measured statistics.

Given these caveats, the weighting schemes, equations (51) and (52), optimally weight galaxies that randomly sample an underlying density distribution, subject to both a selection criteria in real space and a distortion that shifts the radial coordinate of each galaxy. The weights correct statistically for the distortion, selection and shift when measuring a specific statistic of the density field.

We shall return to the effects of these these weighting schemes on the transition matrices later.

## 6   APPLICATIONS TO SIMULATED AND REAL DATA

### 6.1   N-body simulations

We have tested the method using a random subsample of an N-body simulation provided by Bertschinger & Gelb (1991 for details). We choose an observer at random, and form the observer's redshift space map within a large sphere. In order to make the sphere radius large compared to the scale at which the fluctuations become nonlinear, we repeat the box before constructing the redshift space map. In the simulation we used, the power spectrum in real space is rather accurately given by $P(k) = (k/k_0)^{1.55}/(2\pi^2 k^3)$, with $k_0 = 0.45h$ Mpc$^{-1}$. In the first test, we analyse the real-space map, constructing likelihood contours for the quantities $\beta$ and $k_0$. In essence the real-space map has $\beta = 0$. The results are shown in Figure 4, which shows the maximum likelihood solution to be an unbiased measure of both $k_0$ and $\beta$. In Figure 5 we analyse the redshift space map, constraining the power spectrum to have the right form. The maximum likelihood solution lies close to the expected $\beta = 1$.

### 6.2   IRAS 1.2 Jy sample

We have applies the method to the IRAS 1.2 Jy sample, containing $\sim 5000$ galaxies. We form a catalogue with $r_{max} = 200 h^{-1}$ Mpc and a conservative galactic plane cut of $|b| > 10°$, to minimise any residual confusion with galactic sources. This leaves 4511 galaxies, with a mean observed density (Fisher, private communication) of

$$\rho_0 = a r^{-2\alpha} (1 + (r/r_*)^2)^{-\beta_*} \tag{53}$$

with $a = 0.189$, $\alpha = 0.421$, $\beta_* = 1.913$ and $r_* = 50.1$. Units are $h^{-1}$ Mpc. We assume a parameterised form for the power spectrum given by Peacock & Dodds (1994):

$$P(k) = \frac{2\pi^2 A (k/k_0)^{1.5}}{k^3 [1 + (k/k_c)^{-2.5}]} \tag{54}$$

where we take the amplitude $A$ to be a free parameter. $k_0 = 0.29$, $k_c = 0.039h$ Mpc$^{-1}$. $A$ is related to the variance in the real-space *number* overdensity in spheres of radius $8h^{-1}$ Mpc by $\sigma_{8,IRAS} = 0.69 \sqrt{A}$. (Note that $\sigma_8$ is often a linear theory extrapolation; the value here is calculated from the assumed present-day power spectrum).

We treat $\beta$ and $\sigma_{8,IRAS}$ as free parameters, and present results including modes up to $k = 0.1h$ Mpc$^{-1}$. This gives a total of 604 modes, with maximum values $l = 17$ and $n = 6$. The convolutions contain modes up to $l = 30$ and $n = 20$. Shot noise is not very significant for these modes, with signal-to-noise rarely less than, and usually far in excess of, ten, even for the highest $l$ modes considered.

Likelihood contours are shown in the $\sigma_{8,IRAS} - \beta$ plane, in Figure 6 for the weighting scheme designed to minimise variance in the measured power. It is impractical to choose a different weighting scheme for each wavenumber, so we take a single power equal to 6000 $h^{-3}$ Mpc$^3$, roughly what we expect at a wavenumber of 0.15 $h$ Mpc$^{-1}$. The maximum likelihood values are $\beta = 1.1$ and $\sigma_{8,IRAS} = 0.68$, in agreement with the results of Fisher et al. 1994a. Note that this graph also contains the redshift space r.m.s., from the limit $\beta \to 0$, $\sigma_{8,IRAS,z} = 0.87 \pm 0.05$, as found by Feldman et al. 1994 for the QDOT sample, and marginally higher than the value of 0.80 found by Fisher et al. 1993 for the 1.2-Jy sample. The error analysis is complicated because the likelihood contours are very elongated. The ridge of likelihood extending towards high $\beta$ arises from degeneracy between $\beta$ and the amplitude of the power spectrum, which arises if the redshift space power is much larger than the real-space power because of large distortion, as is clear from examination of (27) in the high-$\beta$ limit.

The formal error from $\delta \ln \mathcal{L} = -0.5$ projected on to the parameter axes gives the formal errors:

$$\beta = 1.1 \pm 0.3 \tag{55}$$
$$\sigma_{8,IRAS} = 0.68 \pm 0.05. \tag{56}$$

The errors are perhaps unduly optimistic. The ridge of likelihood extending towards $\sigma_{8,IRAS} = 0.87$ does not exclude $\beta = 0$ with confidence, the 95% confidence level being at $\delta \ln \mathcal{L} = -3$ for the Gaussian approximation (Press et al. 1992). If we were to increase the wavenumber range so that nonlinear effects became important, we would expect the maximum likelihood value of $\beta$ to decrease. We begin to see this when we increase $k_{max}$ to 0.12; the maximum likelihood value drops to 0.5, and the results become considerably noisier than those in Fig. 6.

One of the powerful aspects of this technique is that we have control over the wavenumbers involved, and this wavenumber limit allows us to avoid nonlinearities. Peculiar velocities of the order of 300 km s$^{-1}$ will affect wavenumbers down to $\sim 0.3$, which is also roughly where the density field becomes nonlinear. For this reason we expect the correction for wavenumbers less than 0.1 to be small, and this is indeed the case. Figure 6(b) includes the correction, 6(a) does not.

### 6.3 The value of $\Omega_0$, with COBE, inflation and the abundance of rich clusters

An interesting results is obtained if we use measurements of the microwave background made with the Cosmic Background Explorer (COBE) satellite (Smoot et al. 1992), and make the further assumptions, motivated by inflation, that the power spectrum $P(k) \propto k$ as $k \to 0$, and that the Universe is spatially flat, which requires a cosmological constant if $\Omega_0 < 1$. The $\Omega_0^{0.6}$ dependence of the velocities is virtually unchanged by the presence of a cosmological constant (Lahav et al. 1991) and gravitational instability theory allows us to relate the amplitude of the power spectrum to the COBE temperature fluctuations (Efstathiou et al. 1992):

$$P(k) \to Bk; B = \frac{6\pi^2}{5}\left(\frac{Q_{rms}}{T}\right)^2\left(\frac{2c}{H_0}\right)^4 \Omega_0^{-1.54}\left(\frac{h}{\text{Mpc}}\right)^4 \quad (57)$$

With a quadrupole of $17.0 \times 10^{-6}$ K (Banday et al. 1994) and $T = 2.735$K, we get

$$P(k) = 5.93 \times 10^5 k \Omega_0^{-1.54}\left(\frac{Q_{rms}}{1.7 \times 10^{-5}}\right)^2 (h^{-1}\text{Mpc})^3 \quad (58)$$

with $k$ measured in $h$ Mpc$^{-1}$. Relating this to our parameterised power spectrum, we find the bias parameter, assumed independent of scale, at least on these large scales:

$$b = 0.842\sqrt{A}\Omega_0^{0.77} = 1.22\sigma_{8,IRAS}\Omega_0^{0.77} \quad (59)$$

We can also incorporate the constraint found by White et al. (1993) from the abundance of rich clusters:

$$\sigma_{8,\rho} \simeq 0.57\,\Omega_0^{0.6} \quad (60)$$

with a quoted error of about 10%. We can combine these data to provide weak constraints on the parameters $\Omega_0$ and $b$. This will be the subject of a forthcoming paper (Taylor & Heavens, in preparation), but preliminary results are that $\Omega_0 \gtrsim 0.2$ and $b \gtrsim 0.35$.

### 6.4 Peculiar velocity power spectrum

Although we have assumed gravitational instability for most of this analysis, it is not strictly necessary. Under less restrictive assumptions, we can calculate the peculiar velocity power spectrum required to make the observed density data consistent with an underlying isotropic field. We need to make the following assumptions explicitly: the velocity field is curl-free; galaxies obey the continuity equation; modes grow independently of $k$.

In this case, we may write the peculiar velocity in terms of a velocity potential, $\mathbf{v}(\mathbf{r}) = -\nabla\Psi$, and the linear continuity equation in expanding coordinates gives, $\nabla^2\Psi = a(t)\partial\delta/\partial t$. Assuming modes grow independently of $k$ gives $a(t)\partial\delta/\partial t = F(t)\delta$. The power spectrum for the velocity is then $\langle|\mathbf{v_k}|^2\rangle = F(t)^2\langle|\delta_\mathbf{k}|^2\rangle/k^2$. For gravitational instability, $F(t)$ is just $\beta/H_0$, so by determining $\beta$ and $\langle|\delta_\mathbf{k}|^2\rangle$ by likelihood techniques we determine the velocity power spectrum, independently of the assumption of gravitational instability:

$$\Delta_v^2 = 100^2 \frac{(A\beta^2)_{max}(k/k_0)^{1.5}}{k^2[1+(k/k_c)^{-2.5}]}\,(\text{kms}^{-1})^2 \quad (61)$$

defined such that the variance in the three-dimensional peculiar velocity field is $\sigma_v^2 = \int \Delta_v^2 d\ln k$. $k_0 = 0.29$ and $k_c = 0.039 h$ Mpc$^{-1}$ are obtained from the assumed power spectrum shape. Since $(A\beta^2)_{max} \simeq 1$, we obtain the required velocity power spectrum, shown in Figure 7. Note, that, although the expression (61) involves $\beta$, the calculated $\Delta_v^2$ is independent of the assumption of gravitational instability.

## 7 DISCUSSION AND FUTURE PROSPECTS

We have presented here a very powerful method for extracting the quantity $\beta \equiv \Omega_0^{0.6}/b$ and the power spectrum from near all-sky galaxy catalogues with the minimum of assumptions. In essence, the degree of anisotropy in the redshift space catalogue is exploited to measure the distortion from the expected real-space isotropy. The method is exact to first-order, so represents a significant advance over Fourier methods. In addition, as a power spectrum approach, we can limit our application easily to scales in which the linear approximation holds, one of the reasons why it is superior to correlation function studies. The method exploits the presence of peculiar velocities, but does not rely on actual measurements of peculiar velocities, and thus is not sensitive to any uncertainties over how well they can be measured. We believe that this formalism represents the best method for analysing near all-sky galaxy redshift catalogues.

It is worth comparing the methods presented here and in Fisher et al. 1994a. In the latter case, the galaxies are weighted with arbitrary functions of redshift, and projected on to the sky. The resulting sky distributions are then analysed with spherical harmonics. This has an advantage over the method of this paper in that one can try to tailor the weighting functions to maximise the effect of redshift distortions. In one sense, the work in this paper can be regarded as a particular choice for the set of weighting functions, but this misses the power of the method. The expansion in spherical harmonics is analogous to a full Fourier expansion in cartesian coordinates, offering similar advantages such as direct measurement of the power spectrum (albeit convolved with some window functions) and direct analysis of anisotropy of the coefficients. An obvious extension to the application discussed here is to allow the form of the power spectrum to be free. This work is in progress, and should allow a simultaneous determination of $P(k)$ and $\beta$. A further application of this technique is to surveys which are not all-sky. Without all-sky coverage, there is mixing of different modes with different $l$ and $m$. This makes the computation of likelihoods computationally expensive, but the complications can be minimised if the survey has azimuthal symmetry. In this case mixing only occurs between modes of the same $m$ value, and the resulting likelihood matrices become block-diagonal. This is an important simplification which makes the computations tractable.

Our analysis of the IRAS 1.2 Jy galaxy catalogue yields the following parameters: $\beta \simeq 1.1 \pm 0.3$, and a real-space amplitude $\sigma_{8,IRAS} \simeq 0.68 \pm 0.05$, confirming values found by other methods (e.g. Fisher et al. 1994a, Peacock & Dodds 1994, Dekel et al. 1993). However, with the exception of Fisher et al. 1994a, the other methods which do not use the peculiar velocity field directly have some further approximations which are hard to justify, or use correlation function distortions, which are hard to separate into linear and nonlinear regimes, and which also suffer from small-angle approximations breaking down.

## ACKNOWLEDGMENTS

ANT is supported by a SERC research assistantship. We acknowledge useful discussions with Ofer Lahav, Karl Fisher, Donald Lynden-Bell, Caleb Scharf, John Peacock, Bill Ballinger and Jerry Ostriker. We are grateful to Marc Davis and the rest of the 1.2 Jy IRAS team for providing the survey prior to publication. The software was written using STARLINK facilities.

## APPENDIX A: Radial and Transverse Velocity Fields

Given the spherical nature of the problem of redshift distortions in an all-sky galaxy survey, it is natural to express the velocity field in terms of a spherical vector basis. This allows us to set the boundary conditions on the surface of the enclosing sphere in an uncomplicated way that avoids mixing the 3 cartesian components, and separates out the radial components from the transverse components of the velocity field for us. In general, we can also decompose the velocity field into potential and vortical flows. However, on the scales that interest us the generation of vortical flows by baroclinic interactions and orbit crossing are negligible.

The velocity field may be split into radial and transverse components by

$$\mathbf{v} = u\hat{\mathbf{r}} - \frac{\hat{\mathbf{r}} \times \mathbf{J}}{r}, \qquad (62)$$

where $u(\mathbf{r}) = \hat{\mathbf{r}}.\mathbf{v}$ is the radial velocity part and $\mathbf{J} = \mathbf{r} \times \mathbf{v}$ is the transverse term. Given that the velocity field is potential, we can define a scalar field, $\Psi(\mathbf{r})$, by

$$\begin{aligned}\mathbf{v} &= \nabla\Psi, \\ &= \hat{\mathbf{r}}(\hat{\mathbf{r}}.\nabla)\Psi - \frac{i}{r}(\hat{\mathbf{r}} \times \mathbf{L})\Psi,\end{aligned} \qquad (63)$$

where on the second line we have performed a similar decomposition into radial and transverse components of the gradient operator, and we have defined $\mathbf{L} \equiv -i\mathbf{r} \times \nabla$ as the classical orbital angular momentum operator.

Expanding $\Phi(\mathbf{r})$ into spherical harmonics we find

$$\mathbf{v} = \sum_{lmn} c_{ln} \Psi_{lmn} \times \left( j'_l(k_{ln}r) \mathbf{Y}^{(L)}_{lm}(\Omega) - i \frac{\sqrt{l(l+1)}}{r} j_l(k_{ln}r) \mathbf{Y}^{(M)}_{lm}(\Omega) \right), \quad (64)$$

where we have introduced the orthonormal vector spherical harmonics $\mathbf{Y}^{(L)}_{lm}$ and $\mathbf{Y}^{(M)}_{lm}$. These are defined in terms of the scalar spherical harmonics, $Y_{lm}$ and obey the same eigenfunction conditions;

$$L^2 \mathbf{Y}_{lm} = l(l+1) \mathbf{Y}_{lm}, \quad L_z \mathbf{Y}_{lm} = m \mathbf{Y}_{lm}. \quad (65)$$

It is fairly straightforward to show that these can only be satisfied by the vectors

$$\begin{aligned} \mathbf{Y}^{(L)}_{lm} &= \hat{\mathbf{r}} Y_{lm}, \\ \mathbf{Y}^{(E)}_{lm} &= \frac{1}{\sqrt{l(l+1)}} \mathbf{L} Y_{lm}, \\ \mathbf{Y}^{(M)}_{lm} &= \hat{\mathbf{r}} \times \mathbf{Y}^{(E)}_{lm}, \end{aligned} \quad (66)$$

where the factor of $1/\sqrt{l(l+1)}$ is required for orthonormality. These orthonormality relations are given in Appendix C.

**APPENDIX B**: Effects of small–scale velocity structures.

Here we wish to show that the effect of small scale virialized clusters in redshift space can be expressed by the relation

$$D'_{lmn} = S^{mm'}_{ll'nn'} D_{l'm'n'}. \quad (67)$$

Virialized clusters in redshift space have the effect of smearing out the galaxy distribution along the line of sight. In section 4.3 we have modelled this by randomly assigning each galaxy a displaced position, which is then allowed to vary like a Gaussian distributed variable. We therefore want to relate this picture to the above expression. Perturbing the positions of galaxies in the survey by the random displacement $\varepsilon(\mathbf{r})$, and averaging over all realisations we find

$$\begin{aligned} D'_{lmn} &= \frac{c_{ln}}{V} \langle \int \rho_0(\mathbf{r}) \delta(\mathbf{r}) j_l(k_{ln}s') Y^m_l(\Omega_{s'}) d^3r \rangle_\varepsilon, \\ &= \frac{c_{ln}}{V} \int \rho_0(\mathbf{r}) \delta(\mathbf{r}) \int p(\mathbf{s}-\mathbf{s}') j_l(k_{ln}s') Y^m_l(\Omega_{s'}) d^3s' d^3r, \\ &= \frac{c_{ln}}{V} \int \rho_0(\mathbf{r}) \delta(\mathbf{r}) \sum_{l'n'm'} S^{mm'}_{ll'nn'} j_{l'}(k_{l'n'}s) Y^{m'}_{l'}(\Omega_s) d^3r, \\ &= \sum_{l'n'm'} S^{mm'}_{ll'nn'} D_{l'm'n'}, \end{aligned} \quad (68)$$

where we have expanded the distribution function of the $\varepsilon(\mathbf{r})$'s as in equation (35).

**APPENDIX C**: Orthogonality relations

When preforming calculations in spherical harmonics, it is useful to have reference to the orthogonality, or closure, relations harmonic functions obey. In all cases these can be expressed succinctly in the notation of Dirac as

$$\sum_{lmn} \langle \mathbf{x} | lmn \rangle \langle lmx | \mathbf{x}' \rangle = \delta_D(\mathbf{x} - \mathbf{x}') \quad (69)$$

and

$$\int \langle lmx | \mathbf{x} \rangle \langle \mathbf{x} | l'm'n' \rangle d^3x = \delta^K_{ll'} \delta^K_{mm'} \delta^K_{nn'}. \quad (70)$$

However, in actual application, the symmetry's of the problem may require less general notation than that of Dirac. Below we have compiled a list of some of the more frequently used orthonormal relations.

The closure relation for spherical Bessel functions is

$$\int_0^{r_{\max}} j_l(kx) j_l(kx') k^2 dk = \frac{\pi}{2x^2} \delta_D(x - x'), \quad (71)$$

while that for Spherical Harmonics is

$$\int_{4\pi} Y_{lm}(\Omega) Y^*_{l'm'}(\Omega) d\Omega = \delta^K_{ll'} \delta^K_{mm'}. \quad (72)$$

The inverse relations for Spherical Harmonics is given by

$$\sum_m Y_{lm}(\Omega) Y^*_{lm}(\Omega') = \frac{2l+1}{4\pi} P_l(\mu), \quad (73)$$

where $\mu$ is the cosine of the angle subtending $\Omega$ and $\Omega'$. Summation over the remaining $l$ modes yields

$$\sum_{lm} Y_{lm}(\Omega) Y^*_{lm}(\Omega') = \delta_D(\hat{\mathbf{x}} - \hat{\mathbf{x}}'). \quad (74)$$

For systems with azimuthal symmetry these reduce to

$$\int_{-1}^1 P_l(\mu) P_{l'}(\mu) = \frac{2}{2l+1} \delta^K_{ll'} d\mu, \quad (75)$$

and

$$\sum_l (l+1/2) P_l(\mu) P_l(\mu') = \delta_D(\mu - \mu'). \quad (76)$$

Finally, the orthogonality relations for the vector spherical harmonics of any one type are

$$\int_{4\pi} \mathbf{Y}_{lm} \cdot \mathbf{Y}^*_{l'm'} d\Omega = \delta^K_{ll'} \delta^K_{mm'}. \quad (77)$$

**FIGURES**

**Figure 1.** Coefficients of $\beta$ and $\beta^2$. Kaiser's Fourier approach gives coefficients of 2/3 and 1/5. These are recovered in the limit of $k \to \infty$.

**Figure 2.** Convolution $\Phi$ and redshift distortion matrix $V$ elements, showing the mixing of power caused by the selection function in the first case and by redshift distortion in the second. The figures show elements with $l = l'$: $l=5$, $n = 3$ (solid); $l = 10$, $n = 5$ (dashed); $l = 20$, $n = 10$ (dot-dashed). a) Equal weighting, convolutions $\Phi_{nn'}^{ll'}$ (ignoring mask mixing) b) Minimum variance weighting (for power) convolutions, showing narrow range of wavenumbers sampled. c) Redshift distortion matrix elements $V_{nn'}^{ll'}$, for $\beta = 0$ power minimum variance.

**Figure 3.** (a) Selection functions for 1.2 Jy IRAS survey, QDOT and 2 Jy survey (the last is normalised to unity at small separations). (b) weighting functions for 1.2Jy catalogue: dashed curve is minimum variance for the power, assuming $\beta = 0$ (the 'Feldman, Kaiser, Peacock' weighting; dot-dashed and dotted are redshift space corrected.

**Figure 4.** N-body simulation, analysed in *real* space, for which $\beta$ is effectively zero. The power spectrum is accurately approximated by $P(k) = (k/k_0)^{-1.55}$, with $k_0 = 0.45$. The graph shows likelihood contours for $\beta$ and $k_0$, picking out the correct values. Contour levels are separated by $\delta \ln \mathcal{L} = 0.5$. FIGURE AVAILABLE FROM AUTHORS

**Figure 5.** N-body, redshift space. Here we assume the correct power spectrum, and find the likelihood function for $\beta$. The correct value ($\beta = 1$) is within the errors. FIGURE AVAILABLE FROM AUTHORS

**Figure 6.** Likelihood contours for $\beta$ and the normalisation of the real-space number fluctuation spectrum, $\sigma_{8,IRAS}$ for the IRAS 1.2 Jy sample. We assume a parametric form for the power spectrum, allowing the amplitude to vary. The data are weighted with the minimum variance estimator for power, and modes up to a wavenumber $k = 0.1\,h\,\mathrm{Mpc}^{-1}$ are included. Contour levels are separated by $\delta \ln \mathcal{L} = 0.5$. (a) does not include correction for small-scale velocity dispersion, (b) assumes an incoherent three-dimensional velocity dispersion of 300 km s$^{-1}$. On these scales, the correction is very small.

**Figure 7.** Velocity power spectrum, obtained without assuming gravitational instability. The shape is, however, assumed, as in fig. 4.

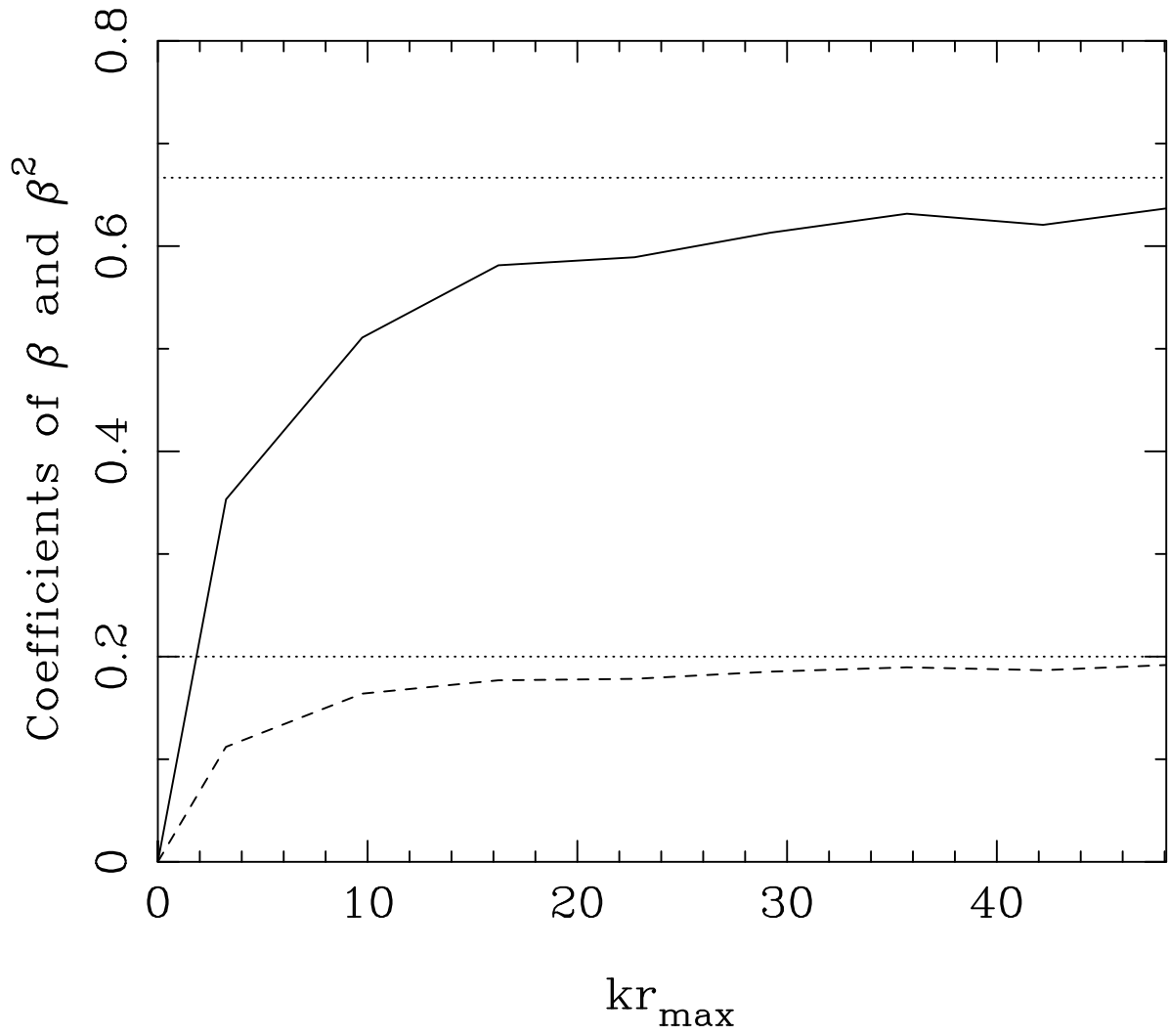

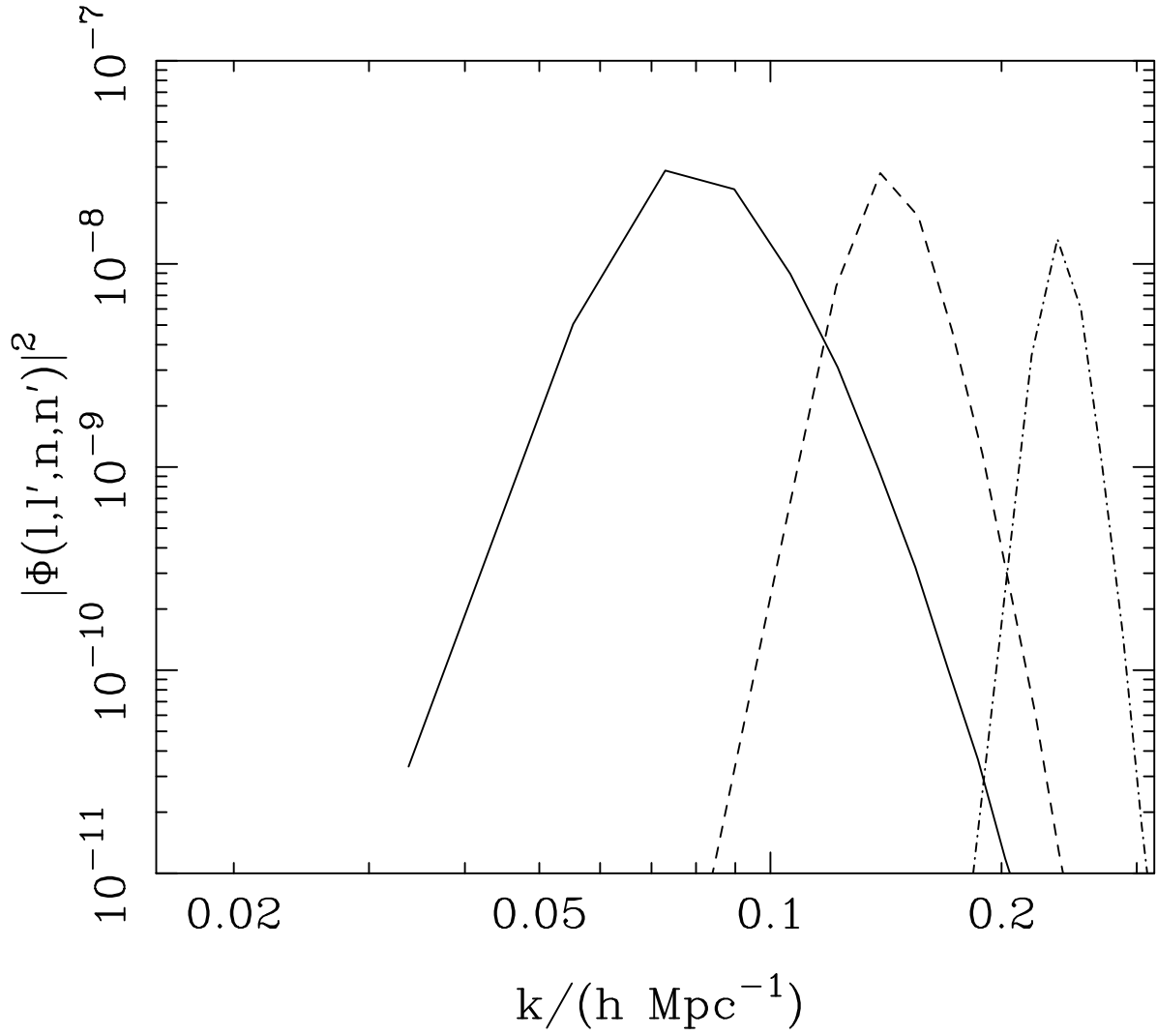

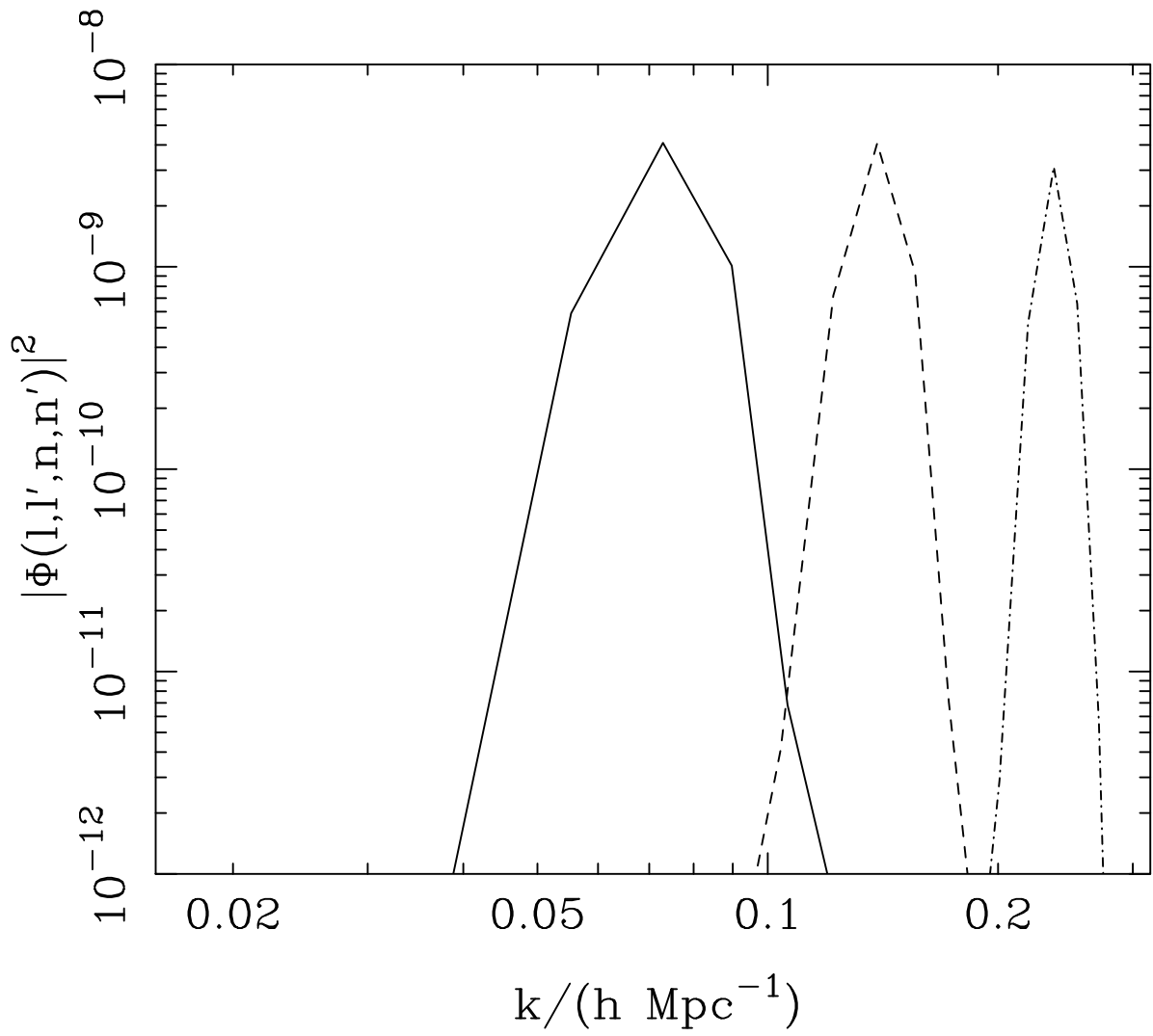

Selection Function Matrices

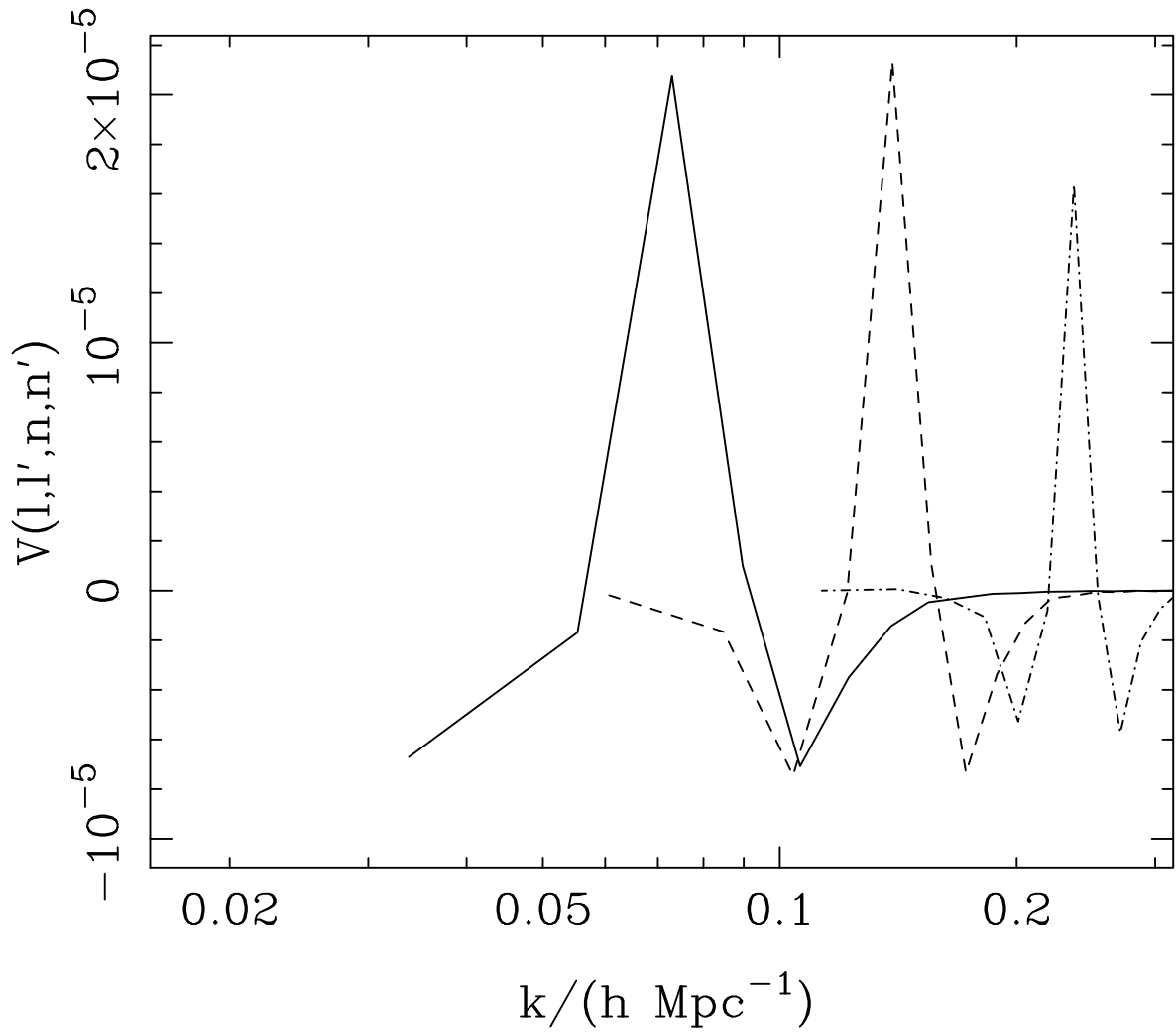

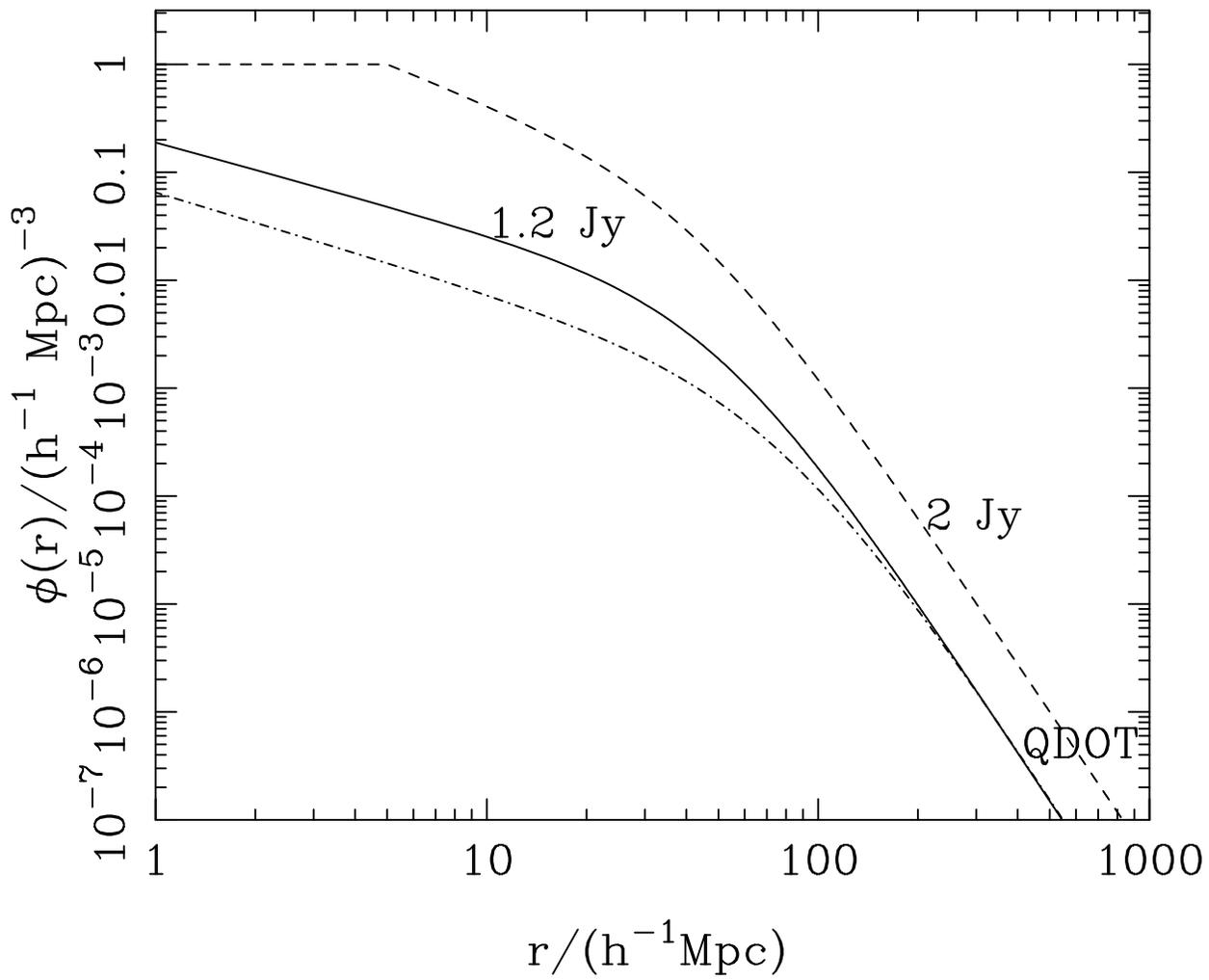

IRAS 1.2Jy, 2Jy and QDOT Selection Functions

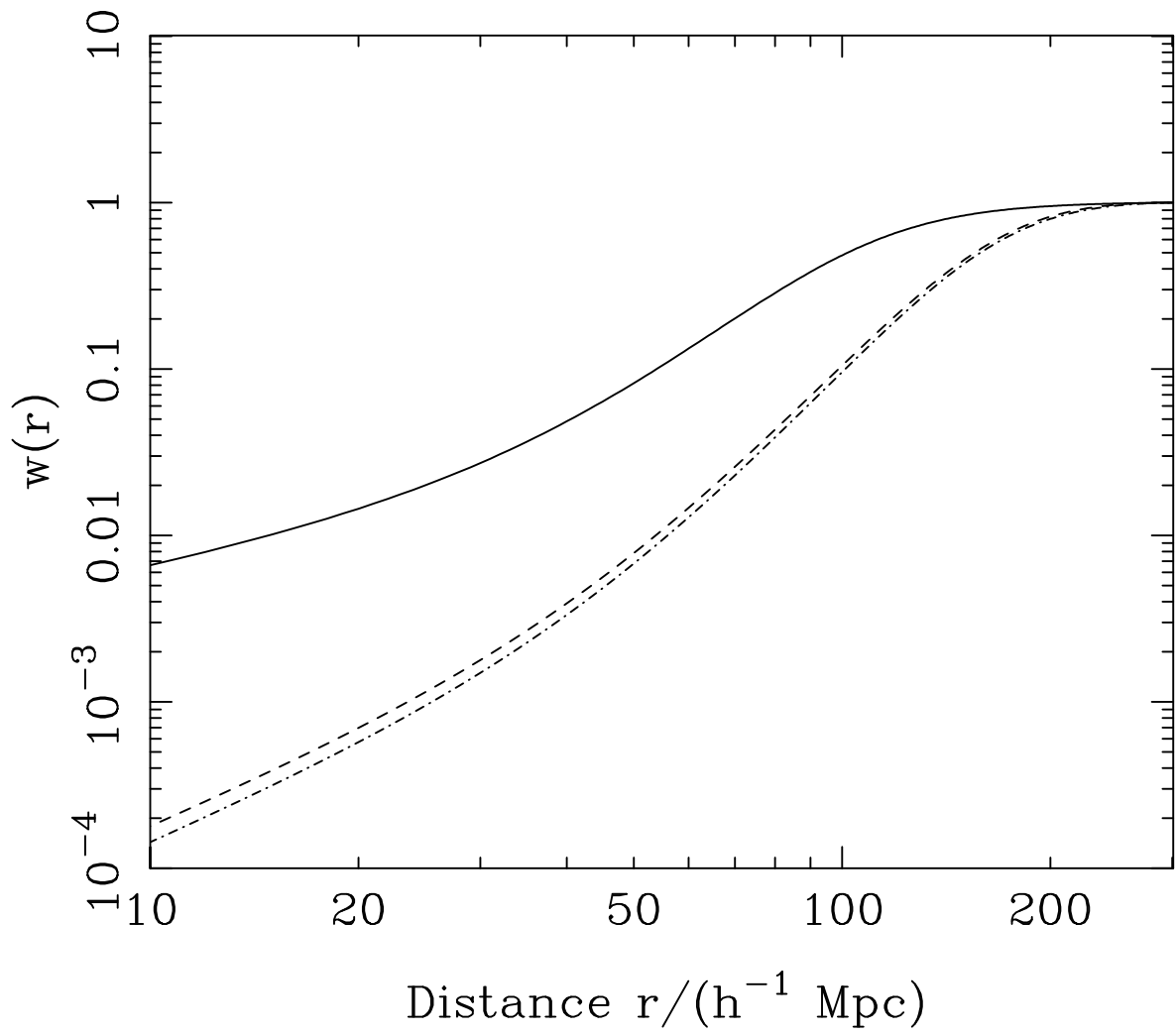

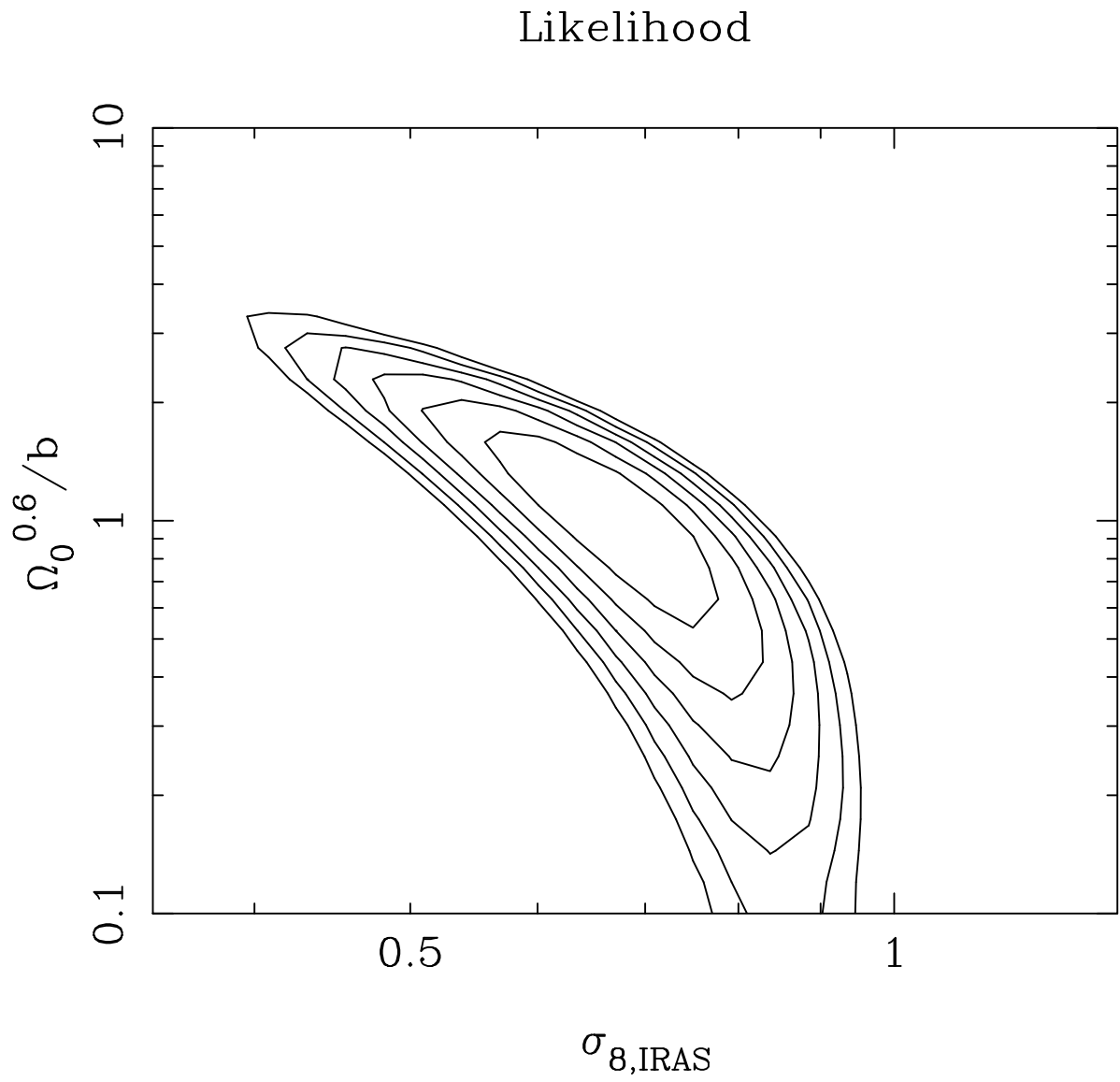

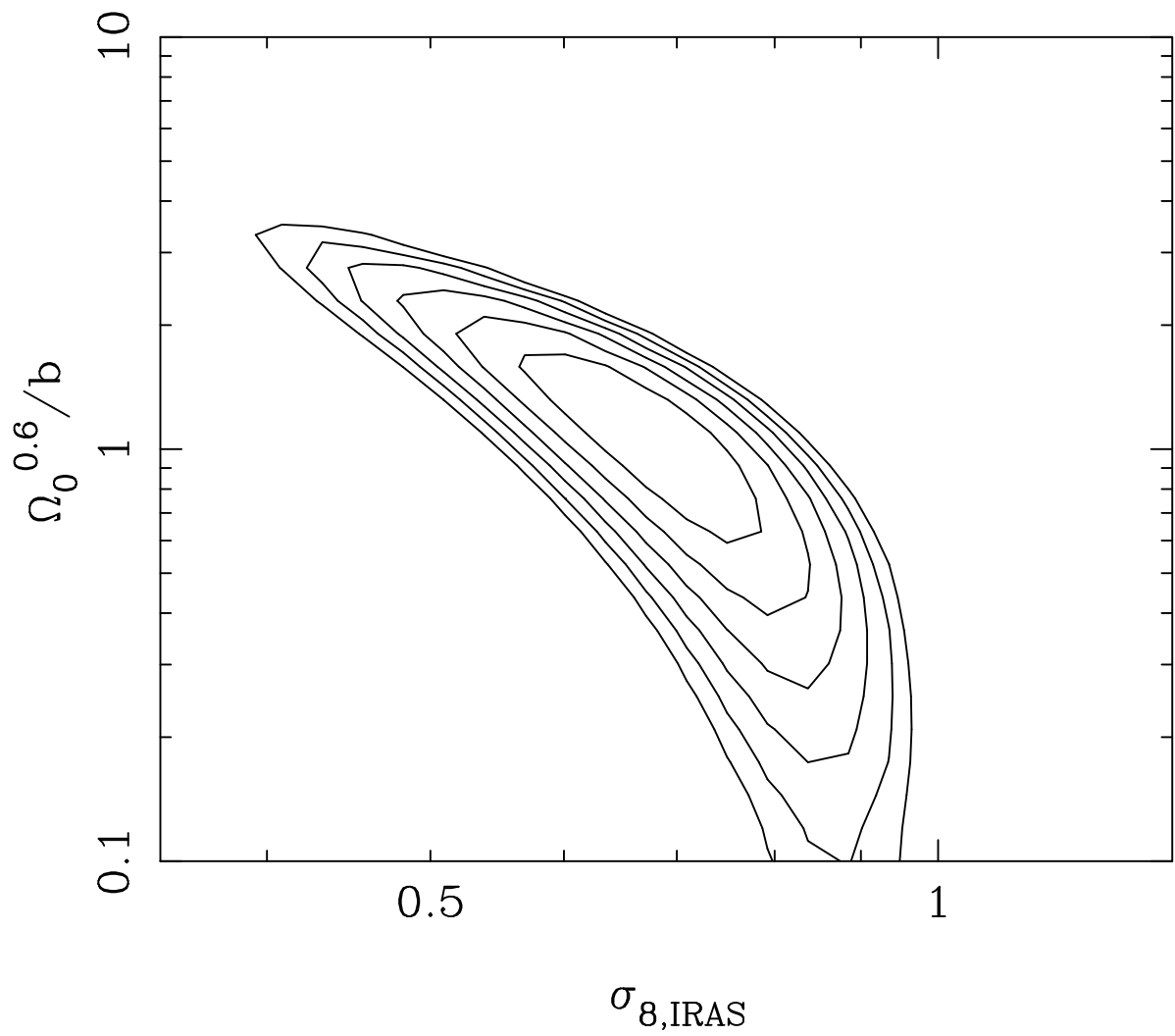

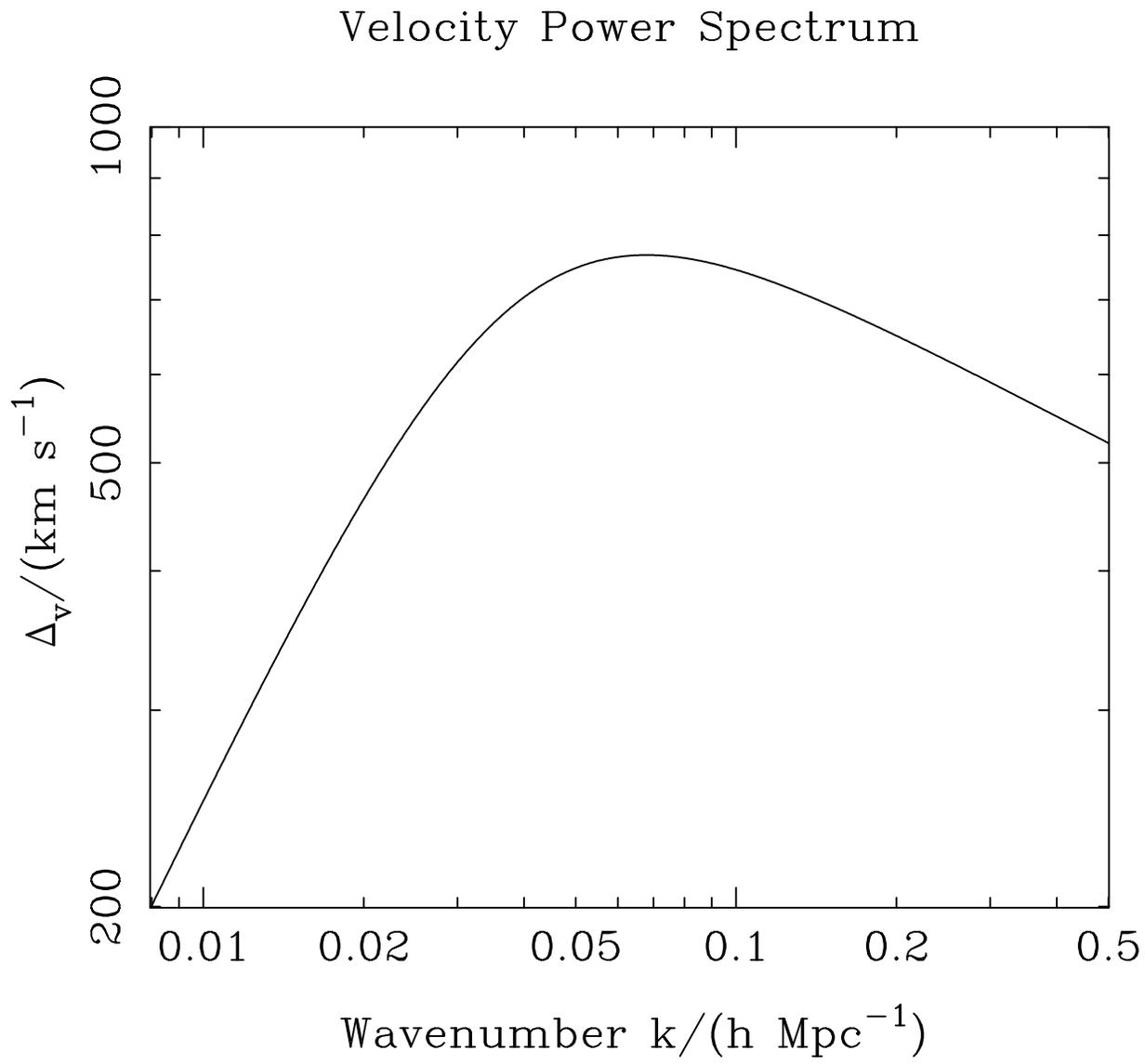